\documentclass[aps,prd,superscriptaddress,nofootinbib,twocolumn]{revtex4-1}

\usepackage{mathtools}
\usepackage{amsfonts}
\usepackage{mathrsfs}

\usepackage{bbm}
\usepackage{slashed}
\usepackage{amsmath}
\usepackage{bm}
\usepackage{tensor}

\usepackage{graphicx}
\usepackage{color}
\usepackage{colortbl}
\usepackage{array}
\usepackage[dvipsnames]{xcolor}
\usepackage{array,multirow,graphicx}

\usepackage{float}
\usepackage{placeins}
\usepackage{booktabs}
\usepackage{subcaption}
\usepackage{makecell}
\usepackage{tabstackengine}

\usepackage{xspace}
\usepackage{siunitx}
\usepackage{hyperref}
\usepackage[nameinlink]{cleveref}
\usepackage{bookmark}

\usepackage{xifthen}
\usepackage{xcolor}
\usepackage{enumitem}
\hypersetup{
	colorlinks,
	linkcolor={red!75!black},
	citecolor={blue!75!black},
	urlcolor={blue!75!black}
}
\usepackage{physics}

\usepackage[utf8]{inputenc}

\setkeys{Gin}{width=0.48\textwidth}

\makeatletter
\newsavebox\myboxA
\newsavebox\myboxB
\newlength\mylenA

\newcommand*\xoverline[2][0.75]{%
	\sbox{\myboxA}{$\m@th#2$}%
	\setbox\myboxB\null
	\ht\myboxB=\ht\myboxA%
	\dp\myboxB=\dp\myboxA%
	\wd\myboxB=#1\wd\myboxA
	\sbox\myboxB{$\m@th\overline{\copy\myboxB}$}
	\setlength\mylenA{\the\wd\myboxA}
	\addtolength\mylenA{-\the\wd\myboxB}%
	\ifdim\wd\myboxB<\wd\myboxA%
	\rlap{\hskip 0.5\mylenA\usebox\myboxB}{\usebox\myboxA}%
	\else
	\hskip -0.5\mylenA\rlap{\usebox\myboxA}{\hskip 0.5\mylenA\usebox\myboxB}%
	\fi}
\makeatother

\newcommand{\imag}{\text{i}}

\graphicspath{{./figures/}}

\usepackage{axodraw2}

\usepackage{xifthen}
\usepackage{xcolor}

\newcommand{\gettitle}{Critical scaling for spectral functions }

\hypersetup{
	colorlinks,
	linkcolor={red!75!black},
	citecolor={blue!75!black},
	urlcolor={blue!75!black}, 
	pdftitle={\gettitle},
	pdfauthor={Kockler},
	pdfkeywords={analytic continuation}
	{correlations functions}
	{functional renormalization group}
	{real time} {spectral function} 
	bookmarksopen=true,
	bookmarksopenlevel=2,
	bookmarksnumbered=true
}

\begin{document}

\title{\gettitle}

\author{Konrad~Kockler}
\affiliation{Institut f\"ur Theoretische Physik, Universit\"at Heidelberg, Philosophenweg 16, 69120
	Heidelberg, Germany}

\author{Jan M.~Pawlowski}
\affiliation{Institut f\"ur Theoretische Physik, Universit\"at Heidelberg, Philosophenweg 16, 69120
	Heidelberg, Germany}
\affiliation{ExtreMe Matter Institute EMMI, GSI, Planckstr. 1, D-64291 Darmstadt, Germany}

\author{Jonas~Wessely}
\affiliation{Institut f\"ur Theoretische Physik, Universit\"at Heidelberg, Philosophenweg 16, 69120
	Heidelberg, Germany}

\begin{abstract}

	We study real-time scalar $\phi^4$-theory in 2+1 dimensions near criticality. Specifically, we compute the single-particle spectral function and that of the $s$-channel four-point function in and outside the scaling regime. The computation is done with the spectral functional Callan-Symanzik equation, which exhibits manifest Lorentz invariance and preserves causality.  
	We extract the scaling exponent $\eta$ from the spectral function and compare our result with that from a Euclidean fixed point analysis.
\end{abstract}

\maketitle

\section{Introduction}
\label{sec:Intro}

We set up a real-time computation of the critical physics of the three-dimensional scalar $\phi^4$-theory. In this regime, the physics is governed by the universality class of the theory, which corresponds to that of the three-dimensional Ising model described by the Wilson-Fisher fixed point. We envisage applying this setup to QCD within the functional renormalisation group (fRG) approach in order to explore the vicinity of its critical end point (CEP) at finite density. In that context, the present real-time approach can be used to study the transport properties of QCD near its CEP.

For the direct real-time access to critical physics, we employ the spectral fRG developed in~\cite{Braun:2022mgx,Fehre:2021eob,Horak:2023hkp}. As a renormalisation group technique, it enables us to study scaling and critical phenomena while simultaneously utilising the Källén-Lehmann representation that provides direct access to correlators in Minkowski space. For recent spectral computations within other functional approaches, see~\cite{Solis:2019fzm, Mezrag:2020iuo, Horak:2020eng, Horak:2021pfr,Horak:2022myj, Horak:2022aza, Pawlowski:2024kxc} for Dyson-Schwinger equations, and~\cite{Eichmann:2023tjk} for Bethe-Salpeter equations.

In this work, we extend the real-time analysis presented in~\cite{Horak:2023hkp} to the scaling regime and solve the system close to the phase transition signalled by a vanishing pole mass. Within sophisticated approximation schemes the fRG approach yields quantitative results for scaling exponents , see, e.g.,~\cite{Litim:2010tt, Balog:2019rrg, DePolsi:2020pjk} and the review~\cite{Dupuis:2020fhh}. These results compares well with those obtained via the conformal bootstrap; see, e.g.,~\cite{El-Showk:2012cjh, El-Showk:2014dwa}. The wealth of results for scaling coefficients derived from Euclidean flows, or more precisely, from fixed point equations, allows us to benchmark the present real-time approach. Accordingly, this regime serves as an ideal testing ground for fRG approximation schemes and regulators. Moreover, further real-time applications of the fRG in a broad variety of research fields provide further benchmarks and technical overlap, see e.g.,~\cite{Gasenzer:2007za, Berges:2008sr, Gasenzer:2010rq, Floerchinger:2011sc, Kamikado:2013sia, Mesterhazy:2013naa, Tripolt:2013jra, Pawlowski:2015mia, Kamikado:2016chk, Jung:2016yxl, Pawlowski:2017gxj, Wang:2017vis, Tripolt:2018jre, Tripolt:2018qvi, Corell:2019jxh, Huelsmann:2020xcy, Jung:2021ipc, Tan:2021zid, Heller:2021wan, Fehre:2021eob, Roth:2021nrd, Roth:2023wbp, Roth:2024rbi, Roth:2024hcu,Topfel:2024iop, Fu:2024rto, Roth:2025hcm}. 

In~\Cref{sec:setup}, we briefly describe the underlying theory as well as the properties of the scaling regime. We introduce the flow equation for the two-point function and present its spectral representation. In~\Cref{sec:Results}, we present results for the spectral function of the propagator in the scaling regime and compare the extracted critical exponents with Euclidean benchmark results. We summarise our findings in~\Cref{sec:Conclusions}.

\section{Spectral flows in the $\phi^4$-theory}
\label{sec:setup}

We investigate the phase transition and the associated real-time dynamics of a 2+1-dimensional scalar $\phi^4$-theory. It is defined by the classical action
\begin{align}
	S[\varphi] & =\int\! \mathrm{d}^3x \,\bigg\{ \frac{1}{2} \varphi\Bigl(-\partial^2 +m_\phi^2 \Bigr)\varphi + \frac{\lambda_\phi}{4!} \varphi^4\bigg\} \,,
\label{eq:S}
\end{align}
with the fundamental field or field operator $\varphi$. The corresponding mean field in a given background (current) is denoted by $\phi=\langle \varphi\rangle$. 
\Cref{eq:S} depends on the classical coupling $ \lambda_\phi $ and the mass $m_\phi^2 $. 
The ratio $ \lambda_\phi/m_\phi$ is the only dimensionless parameter of the theory and determines the strength of correlations in the system. 
This theory exhibits a second order phase transition of the Ising universality class, given by the Wilson-Fisher fixed point of the renormalisation group. It exhibits two independent critical scaling exponents, that are carried by the propagator $G(p)$ of the theory,
\begin{align} 
G(p) (2 \pi)^3\delta(p+q)= \langle \varphi(p) \varphi(q)\rangle_c\,, 
\label{eq:ConTwoPoint}
\end{align}
where the subscript ${}_c$ indicates the connected part. The propagator can be parametrised by
\begin{align}  
G(p) = \frac{1}{Z_\phi(p)\left( p^2 +m_\text{\tiny{pole}}^2\right)}\,,
\label{eq:Gp}
\end{align}
where the mass $m_\text{\tiny{pole}}$ in~\labelcref{eq:Gp} is the pole mass. The wave function $Z_\phi(p)$ carries the non-trivial part of the momentum dependence. 
We shall use an on-shell renormalisation scheme for which the pole mass is simply the classical mass parameter, $m_\phi=m_\text{\tiny{pole}}$, see~\labelcref{eq:onshellrenSymmetric} and the discussion there. 
Within this physical renormalisation scheme the phase transition is approached with $m_\phi^2=m_\text{\tiny{pole}}\to0,$ or equivalently $\lambda_\phi/m_\phi \to \infty$ as discussed in~\cite{Eichmann:2023tjk}. 

At the phase transition the correlation length diverges, $\xi\to \infty$, and the propagator exhibits an algebraic decay or scaling with the critical exponent $\eta$, 
\begin{align}
	G(p)\propto\frac{1}{(p^2)^{1-\eta/2}}\, , \qquad \longrightarrow \qquad Z_\phi(p) \propto (p^2)^{-\eta/2}\,, 
\label{eq:scalingprop}
\end{align}
with the Ising critical exponent $\eta \approx 0.036 $. 
In the present work, we compute this scaling within the real-time setup. To that end we use the spectral Callan-Symanzik (CS) equation put forward in~\cite{Fehre:2021eob, Braun:2022mgx}, which has been implemented for the $\phi^4$-theory in~\cite{Horak:2023hkp}. 
We identify the CS-regulator with the mass parameter of the theory, 
\begin{align}
m_\phi^2=Z_\phi \,k^2\,, 
\label{eq:CSRegulator}
\end{align}
which readily leads to $k=m_\text{\tiny{pole}}$. The parameter $Z_\phi$ is the on-shell wave function of the field, 
\begin{align} 
	Z_\phi=Z_\phi(p^2 = - m_\text{\tiny{pole}})\,.
\end{align} 
This choice is natural for the on-shell renormalisation scheme~\labelcref{eq:onshellrenSymmetric} used in the present work. 
It entails that $k$ is precisely the pole mass and $Z_\phi$ is the wave function of the asymptotic 1-particle state, i.e., the residue of the propagator on the mass pole.

\begin{figure}[t]
	\centering
	\begin{minipage}{0.99\linewidth}
		\includegraphics[width=\textwidth]{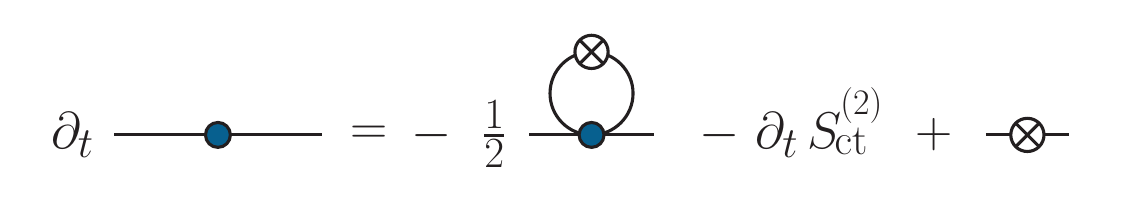}
		\caption{Flow of the inverse two-point function with the scale $k$. Blue vertices denote full 1PI n-point functions and the crossed vertex denotes the derivative of the CS-regulator $\partial_t m^2=Z_\phi k^2 (2-\eta_\phi)$. \hspace*{\fill}  }
		\label{fig:2pointflow}
	\end{minipage}
\end{figure}
%

\subsection{Renormalised Callan-Symanzik equation}
\label{sec:RenCS}

We set up the theory for a sufficiently large mass, $\lambda_\phi/m_\phi \to 0$, where the theory is perturbative. In this limit, the quantum effective action $\Gamma[\phi]$ is approaching the (convex hull of the) classical action, $\Gamma[\phi]\to S[\phi]$. The flow of the quantum effective action with an infinitesimal change of the mass $m_\phi$ towards smaller masses is provided by the renormalised functional Callan-Symanzik equation~\cite{Fehre:2021eob, Braun:2022mgx}, 
\begin{align}
\hspace{-.1cm}	\partial_t \Gamma[\phi] = & \,  \frac{1}{2}\left(2- \eta_\phi\right)Z_\phi k^2\, \text{Tr} \,\Bigl[ \,G[\phi]+\phi^2\Bigr] - \partial_t   S_{\text{ct}}[\phi] \,, 
	\label{eq:CSflow}
\end{align}
where $G[\phi]$ is the full field-dependent propagator 
\begin{align} 
	G[\phi] = \frac{1}{ \Gamma^{(2)}[\phi]}\,, \qquad \Gamma^{(n)}[\phi]= \frac{\delta^n \Gamma[\phi]}{\delta\phi^n} \,. 
\label{eq:Gphi}
\end{align}
The propagator $G(p)$ in~\labelcref{eq:Gp} is obtained by evaluating $G[\phi] $ on the solution $\phi_0$ of the (constant) equation of motion. The (negative) RG-time $t=\text{log}(k/k_{\text{ref}})$ in~\labelcref{eq:CSflow} is measured relative to a suitable reference scale. The anomalous dimension $\eta_\phi$ in~\labelcref{eq:CSflow} is that on the pole, 
\begin{align}
	\eta_\phi= - \frac{\partial_t \, Z_\phi}{Z_\phi}\, .
\label{eq:etaphimu}
\end{align}
The critical exponent $\eta$ at the Wilson-Fisher fixed point is obtained directly from the wave function $Z_\phi(p)$ for $m_\text{\tiny{pole}}=0$, see~\labelcref{eq:scalingprop}. We can also extract it from the zero mass limit of the anomalous dimension $\eta_\phi$ with  
\begin{align}
	\eta=\eta_\phi\left(m_\text{\tiny{pole}} \to 0\right)\,.
	\label{eq:eta_vs_etaphi}
\end{align}
The renormalised CS flow~\labelcref{eq:CSflow} is manifestly finite~\cite{Braun:2022mgx}: it is derived as the CS-limit of UV-regularised finite flows, and all terms with positive UV power-counting dimensions cancel out by further ones that are induced by the change of the UV-regularisation scale. The remainder of this procedure is the flow of the counter-term action $\partial_t S_\text{ct}[\phi]$, which specifies the renormalisation condition for all UV-relevant parameters. In this work, we will make direct use of the real-time nature of our approach and choose an on-shell renormalisation condition for the inverse propagator, 
\begin{align} 
	\Gamma^{(2)}[\phi_0] \Big|_{p^2=-k^2}=0\,, 
\label{eq:onshellrenSymmetric}
\end{align}
evaluated on the equation of motion. For further details on the renormalisation condition both in the symmetric and broken phase see~\cite{Horak:2023hkp}. The phase transition can be approached from both sides, and we choose to resolve the spectral flow in the symmetric phase with $\phi_0=0$ for numerical simplicity. \Cref{eq:onshellrenSymmetric} straightforwardly translates into a condition for the flow equation of the inverse propagator
\begin{align}\nonumber
	\partial_t\Gamma^{(2)}(p) & =-\frac{1}{2}\left(2-\eta_\phi\right)Z_\phi  k^2 D_{\text{tad}}(p)                                                 \\[1ex]
	                             & \hspace{.7cm}+\left(2-\eta_\phi\right)Z_\phi  k^2 - \partial_t S_{\text{ct}}^{(2)}\,.
\label{eq:FlowG2}
\end{align}
With $\phi_0=0$, the diagrammatic part of the flow contains only the tadpole diagram $D_{\text{tad}}(p)$, for a diagrammatic representation see~\Cref{fig:2pointflow}.
The flow of the counter-term action $\partial_t S_{\text{ct}}^{(2)}$ is fixed by the total $t$-derivative of the renormalisation condition~\labelcref{eq:onshellrenSymmetric}, 
\begin{align}
	\partial_t\Gamma^{(2)}\bigg|_{p^2=-k^2}=2Z_\phi k^2\,.
\end{align}
For the final flow equation see~\labelcref{eq:FlowG2Final}.

\subsection{Spectral flows of correlation functions}
\label{sec:specflows}

The derivation in~\Cref{sec:RenCS} holds true for any regulator function, and we have chosen the CS regulator as it preserves two important properties of the theory: Lorentz invariance and causality, see~\cite{Braun:2022mgx}. The third one, ultraviolet finiteness, is guaranteed by the functional renormalisation procedure defined in~\cite{Braun:2022mgx}. This allows us to use spectral representations for propagator and vertices and gives us direct access to real-time physics.

\subsubsection{Two-point function}
\label{sec:TwoPoint}

The flow of the (inverse) two-point function in the full complex frequency plane is obtained by using the Källén-Lehmann representation of the propagator, 
\begin{align}
	G(p) = \int_\lambda\frac{\rho(\lambda)}{\lambda^2+p^2} \,,
\label{eq:specrepprop}
\end{align}
with the spectral function 
\begin{align} 
	\rho(\omega)  =2\Im G\Bigl(p_0\to-\imag(\omega+\imag 0^+), \vec p\,= 0\Bigr)\,.
	\label{eq:rho}
\end{align}
Owing to Lorentz invariance, the propagator is fully determined by its values at vanishing spatial momentum. The spectral function~\labelcref{eq:rho} satisfies the spectral sum rule,
\begin{align} 
\quad 	\int_\lambda \rho(\lambda)=1\,, \qquad 	  \int_\lambda = \int_0^{\infty}\frac{d\lambda^2}{2\pi}\,. 
\label{eq:SpecSum}
\end{align}
The spectral sum rule entails that the total spectral weight is unity, which is in one-to-one correspondence to canonical commutation relations for the field as well as normalised one-particle states. 

The propagator spectral function has the form 
\begin{align}
	\rho(\lambda) & =\frac{2\pi}{Z_\phi}\delta\left(\lambda^2 -m_{\text{\tiny{pole}}}^2\right) +\theta(\lambda-m_{\text{scat}}) \Tilde{\rho}(\lambda) \,.
\label{eq:rhoparam}
\end{align}
It contains a $\delta$-function peak for the one-particle state and a scattering continuum. The onset of the latter in the symmetric phase is given by that of the $1\to 3$ scattering. It is mediated by the four-point scattering vertex, and the momentum dependence of the latter is a crucial ingredient for the spectral flow in the symmetric phase. In the present work, we use an $s$-channel approximation, where the vertex only depends on $s$ with 
\begin{align} 
	s=r^2 \,,\qquad r=p + q\,, 
\end{align}
with the loop momentum $q$ and the incoming momentum $p$ in~\Cref{fig:2pointflow}. This leads to the following parametrisation of the full four-point function,  
\begin{align}
\hspace{-.1cm}	\Gamma^{(4)}(r) =    \lambda_{\phi} + \Gamma^{(4)}_{\text{dyn}}(r)\,,\qquad \Gamma^{(4)}_{\text{dyn}}(r)= \int\limits_\lambda \frac{\rho_{4}(\lambda)}{\lambda^2 + r^2}\,, 
	\label{eq:SplitG4}
\end{align} 
where $\lambda_\phi$ is the momentum-independent part of the vertex. The remaining momentum-dependent part $\Gamma^{(4)}_{\text{dyn}}(p)$ in~\labelcref{eq:SplitG4} admits a spectral representation. Its diagrammatic form and spectral computation is discussed in~\Cref{sec:FourPoint}. Here we 
only remark that the $s$-channel spectral function in~\labelcref{eq:SplitG4} can be extracted similar to that of the propagator by 
\begin{align} 
	\rho_{4}(\omega)=  2 \, \text{Im} \, \Gamma^{(4)}_{\text{dyn}}\Bigl(p_0 = - \imag (\omega + \imag 0^+), \vec p\,= 0\Bigr) \,.	
\label{eq:rho4} 
\end{align}
With the spectral representations of the propagator~\labelcref{eq:specrepprop} and four-point vertex~\labelcref{eq:SplitG4}, the momentum-dependent part of~\labelcref{eq:FlowG2} follows as 
\begin{align}\nonumber
D_{\text{tad}}^{\text{dyn}}(p) =&\, \int_q\,G^2(q)\Gamma^{(4)}_{\text{dyn}}(p+q)\\[1ex]
 =& \int\limits_{\lambda,q}\, \frac{\rho(\lambda_1)\rho(\lambda_2)\rho_4(\lambda_3)}{(\lambda_1^2 + q^2)(\lambda_2^2 + q^2)(\lambda_3^2 + (p+q)^2)} \, ,
\label{eq:Ddyntad}
\end{align}
with 
\begin{align} 
	\int\limits_{\lambda,q} = \int\limits_{\lambda_1}  \int\limits_{\lambda_2} \int\limits_{\lambda_3} \int\limits_q\,, \qquad \int\limits_q = \int \frac{ \textrm{d}^3 q}{(2\pi)^3}\,. 
\label{eq:ints} 
\end{align}
The momentum integration in~\labelcref{eq:Ddyntad} can be performed analytically, and the result is provided in~\labelcref{eq:DdyntadApp} in~\Cref{app:specDiags}. The remaining spectral integrations are performed numerically, and the non-perturbative information is completely carried by the spectral functions. Finally, we use~\labelcref{eq:rho} to extract the spectral function $\rho(\omega)$ from the integrated flow of the inverse propagator~\labelcref{eq:FlowG2}. Importantly, the analytic frequency dependence in~\labelcref{eq:Ipol} allows us to perform the limit $p_0 = - \imag (\omega + \imag 0^+)$ analytically. With the following discussion of the four-point function, this completes the setup for the computation of $\rho(\omega)$.

\subsubsection{Four-point function}
\label{sec:FourPoint}

We close the system of coupled equations for correlation functions with that of the four-point function. We follow~\cite{Horak:2020eng,Braun:2022mgx, Eichmann:2023tjk, Horak:2023hkp} and use the inhomogeneous Bethe-Salpeter equation with a classical scattering kernel. This amounts to a bubble resummation of the four-point function in the $s$-channel,
\begin{align}
	\Gamma^{(4)}(p) = \frac{\lambda_{\phi}}{1+\frac{\lambda_\phi}{2} D_{\textrm{\tiny{fish}}}(p)} \, , 
\label{eq:G4Full} 
\end{align}
and hence the momentum-dependent part $\Gamma_\textrm{dyn}^{(4)}(p)$ of the four-point function~\labelcref{eq:SplitG4} is given by 
\begin{align}
	\Gamma_\textrm{dyn}^{(4)}(p) = \lambda_\phi \frac{  D_{\textrm{\tiny{fish}}}(p)}{\frac{2}{\lambda_\phi}+ D_{\textrm{\tiny{fish}}}(p)} \, . 
	\label{eq:G4Dyn} 
\end{align}
The fish diagram $D_{\textrm{\tiny{fish}}}(p) $ in~\labelcref{eq:G4Full,eq:G4Dyn} is given by  
\begin{align}\nonumber 
D_{\textrm{\tiny{fish}}}(p)   =&\, \int_q \, G(q)G(q+p)\\[2ex]
= &\int_{\lambda,q}\, \frac{ \rho(\lambda_1)\rho(\lambda_2)}{(\lambda_1^2+q^2)(\lambda_2^2+(q+p)^2)}\,.
\label{eq:Dfish}
\end{align}
The computation of the fish-diagram~\labelcref{eq:Dfish} is very similar to the computation of the momentum-dependent part of the tadpole~\labelcref{eq:Ddyntad}. The momentum integration can be performed analytically as well, and the result is provided in~\labelcref{eq:DfishApp} in~\Cref{app:specDiags}. The remaining spectral integrals are integrated numerically. We use~\labelcref{eq:rho4} for the extraction of the spectral function $\rho_4(\omega)$. As for the propagator, the analytic frequency dependence in~\labelcref{eq:tildeIpol} allows us to perform the limit $p_0 = - \imag (\omega + \imag 0^+)$ analytically. This completes the setup for the computation of $\rho_4(\omega)$.

\subsubsection{Wrap-up}
\label{sec:WrapUp}
 
With the split~\labelcref{eq:SplitG4}, the tadpole contribution in~\labelcref{eq:FlowG2} splits into a momentum-independent part and the contribution of $D_{\text{tad}}^{\text{dyn}}(p)$ in~\labelcref{eq:Ddyntad}. The momentum-independent part is cancelled by the counter term $S^{(2)}_\textrm{\tiny{ct}}$ in~\labelcref{eq:FlowG2} due to the renormalisation condition~\labelcref{eq:onshellrenSymmetric}. This condition also eliminates the (flowing) contribution of $D_{\text{tad}}^{\text{dyn}}(p)$ to the pole mass. With these cancellations, we are led to the final form of the flow equation~\labelcref{eq:FlowG2} in the symmetric phase, 
\begin{align}\nonumber
	\partial_t \Gamma^{(2)}(p) =&\, 2 Z_\phi k^2 \\[1ex] 
	&\, \hspace{-1.9cm} -\frac{1}{2}\left(2-\eta_\phi\right)Z_\phi k^2\,\left[  D_{\text{tad}}^{\text{dyn}}(p)-   D_{\text{tad}}^{\text{dyn}}(p^2=-k^2)\right]  \,.   
\label{eq:FlowG2Final}
\end{align}
The first term on the right-hand side of~\labelcref{eq:FlowG2Final} comprises the tree-level flow of the pole mass $m_\text{\tiny{pole}}
=k$. The term in the second line includes the momentum-dependent quantum fluctuations that are induced by the change of the pole mass. The current approximation takes into account a momentum-dependent four-point function, but the respective effective potential 
\begin{align}
V_\textrm{eff}(\rho) = \frac{\Gamma[\rho]}{{\cal V}_3}\,,\qquad \rho = \frac{\phi^2}{2}\,,
\label{eq:Veff} 
\end{align} 
does not include higher order couplings, i.e., ${V_\textrm{eff}^{(n>2)}(0)\equiv 0}$ with ${V_\textrm{eff}^{(n)} = \partial_\rho^n V_\textrm{eff}}$. We consider constant $\rho$ in \labelcref{eq:Veff}, and ${{\cal V}_3 = \int_x}$ is the three-dimensional volume. Moreover, we have 
\begin{align} 
	V_\textrm{eff}^{(2)}(0) = \Gamma^{(4)}(p=0)= \frac{\lambda_\phi}{1 +\frac{\lambda_\phi}{2} D_{\textrm{\tiny{fish}}}(0)}\,.
\end{align}
We close this Section with the remark, that the spectral fRG also allows for the inclusion of a full effective potential. Such an approximation takes into account scatterings to all orders. This includes implicitly also t- and u-channel contributions, and will be the subject of future work. We outline the procedure towards the inclusion of the full effective potential in~\Cref{app:fullpotential}.

\begin{figure*}
	\centering
	\begin{minipage}{0.99\linewidth}
		\centering
		\begin{subfigure}[t]{0.48\linewidth}
			\includegraphics[width=\textwidth]{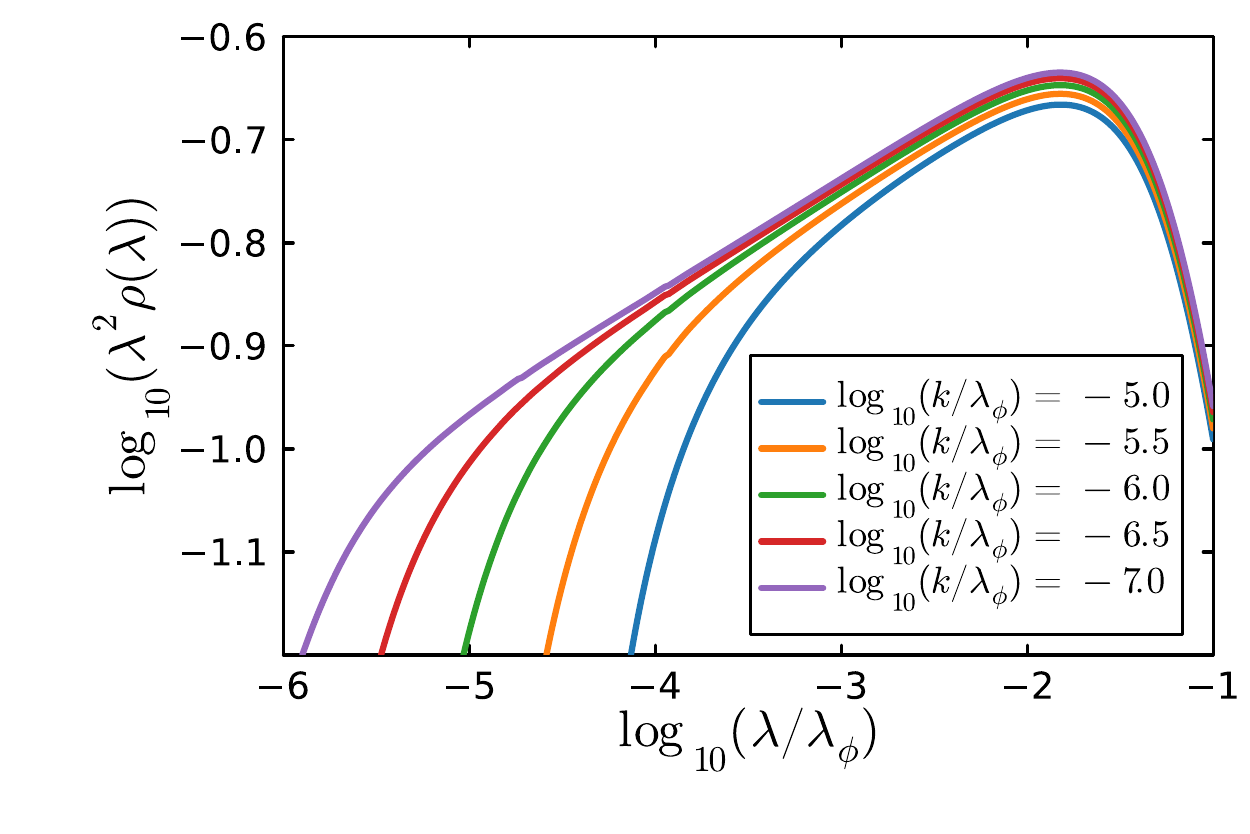}
			\caption{Scaling behaviour of the propagator spectral function: A clear scaling regime in which the propagator spectral function exhibits a power law behaviour is visible and extends to the IR for decreasing masses. The multiplication with $\lambda^2$ removes the trivial dimensional scaling. \hspace*{\fill}}
			\label{fig:rho}
		\end{subfigure}\hfill
		\begin{subfigure}[t]{0.48\linewidth}
			\includegraphics[width=\textwidth]{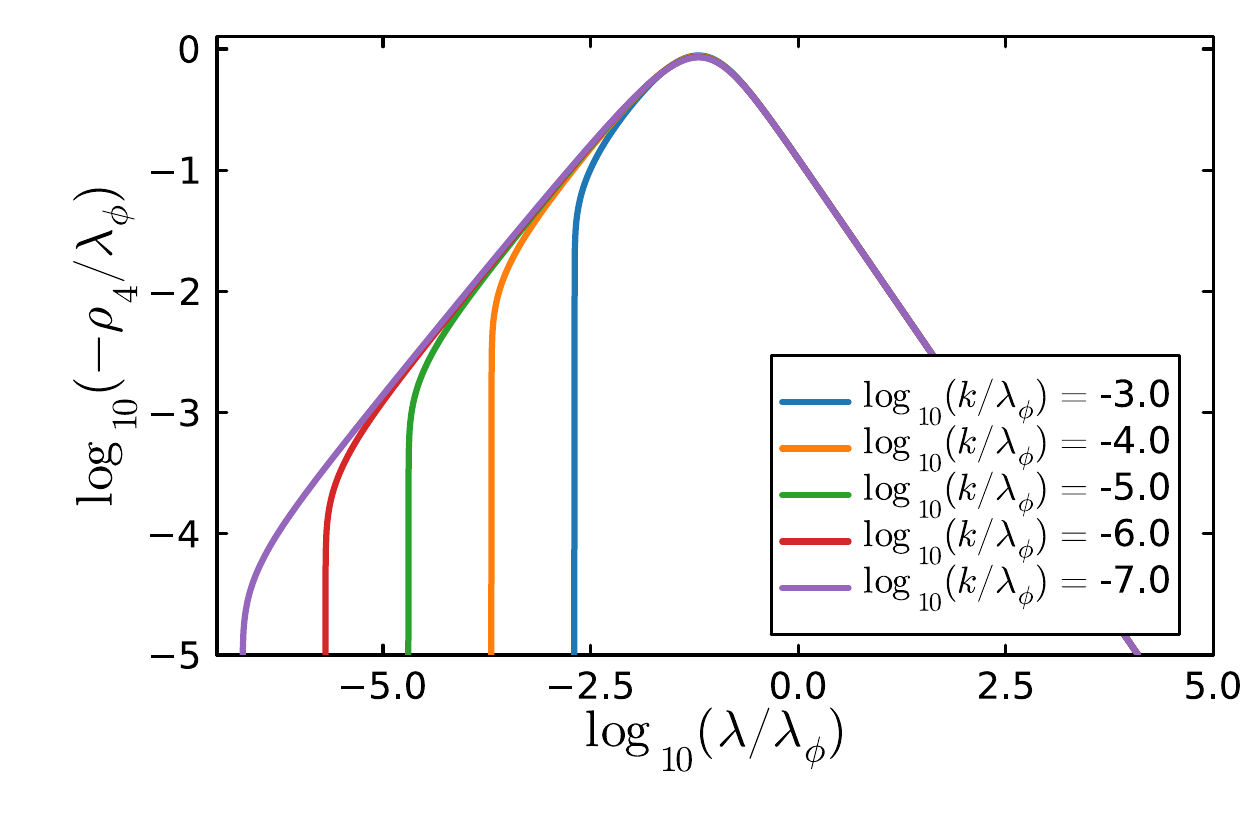}
			\caption{Scaling behaviour of the vertex spectral function: Both the trivial UV and the non-trivial IR scaling regimes as well as the onsets at $m_{\text{\tiny{scat}}}=2m_{\text{\tiny{pole}}}$ are visible. \hspace*{\fill}}
			\label{fig:rho4}
		\end{subfigure}
		\caption{Propagator and vertex spectral function in the scaling regime: The different curves correspond to spectral functions at different pole masses $m_\text{\tiny{pole}}=k$, varied over two and seven orders of magnitude respectively. We measure the spectral parameter in units of the classical coupling $\lambda_\phi$.\hspace*{\fill} } 
		\label{fig:rho+rho4}
	\end{minipage}
\end{figure*}
%

\section{Results}
\label{sec:Results}

In this Section, we discuss our results for the spectral functions of the propagator and the four-point function. Apart from obtaining the full momentum dependence of these correlation functions, we also extract the critical exponent $\eta$ from our computations in three different ways: (1) the scaling behaviour of the spectral function of the propagator (\Cref{sec:SpectralScaling}), (2) the scaling behaviour of the four-point function (\Cref{sec:SpectralScaling} and~\Cref{app:rho4}), and (3) the cutoff scaling of the wave function on the pole (\Cref{sec:flowofZ}).  

We start with a brief survey over commonly used approximation schemes for critical physics and beyond. A standard  approximation scheme used for the analysis of critical phenomena in scalar theories is the derivative expansion, see 
~e.g.,~\cite{Litim:2010tt, Balog:2019rrg, DePolsi:2020pjk} and the review \cite{Dupuis:2020fhh}. In the lowest order, the zeroth order or local potential approximation (LPA), only momentum-independent quantum corrections are taken into account. Then, successively higher momentum orders are incorporated within an expansion in $p^2/k^2$. At higher orders, including the fourth and in particular the sixth order, this expansion leads to quantitative precision. In turn, full momentum dependences of correlation functions are commonly used in QCD as they carry the information on confinement, as well as a crucial dynamical part of the chiral phase transition. This starts with \cite{Ellwanger:1995qf, Ellwanger:1996wy}, see also \cite{Bergerhoff:1997cv, Pawlowski:2003hq, Mitter:2014wpa, Cyrol:2017ewj, Ihssen:2024miv, Fu:2025hcm}, see also the review \cite{Dupuis:2020fhh} and \cite{fQCD}. In scalar models, momentum dependences can be incorporated with the BMW-approximation \cite{Blaizot:2005xy, Blaizot:2005wd, Blaizot:2006vr}, for applications see \cite{Dupuis:2020fhh}. The BMW-approximation requires the computation of the two-dimensional dressing function of the two-point function in terms of momenta $p$ and constant fields. A further approximation scheme with full momentum and field dependence and qualitatively reduced computational costs is provided by the scheme that underlies parts of the present work and has been put forward in \cite{Helmboldt:2014iya}. There it was also shown in a Yukawa model, that pole masses and further observables are already captured well on the percent level within a simple step beyond LPA by including a cutoff-dependent but momentum-independent wave function, the LPA${}^\prime$ approximation. Note however that in the present scalar model this can only potentially work well in the broken regime. In the broken phase this approximation lacks any momentum dependence for the propagator and has a trivial wave function $Z_\phi=1$.  

In conclusion, the present approximation carries the minimal necessary ingredients for capturing the full dynamics. We present a comparison between the results from fully momentum-dependent spectral propagators in this work and from LPA${}^\prime$ in different truncations in~\Cref{sec:LPAprime}. Further details are provided  in~\Cref{app:LPAbenchmark}.

\subsection{Critical scaling}
\label{sec:scaling specfuncs}

The scaling form of the propagator~\labelcref{eq:scalingprop} carries over to the single-particle spectral function, 
\begin{subequations} 
\label{eq:Scalingrhos}
\begin{align}
	\rho(\lambda)\propto \lambda^{-2+\eta}\,. 
\label{eq:scalingrho}
\end{align}
The exponent in~\labelcref{eq:scalingrho} is the critical exponent $\eta \approx 0.036$ of the three-dimensional Ising model, if no approximation is applied. The $s$-channel spectral function $\rho_4$ in~\labelcref{eq:rho4} of the four-point scattering vertex shows the scaling
\begin{align}
	\rho_4(\lambda) \propto \lambda^{1-2\eta}\, .
	\label{eq:scalingrho4}
\end{align}
\end{subequations}
The scaling~\labelcref{eq:scalingrho4} follows readily from~\labelcref{eq:SplitG4,eq:G4Dyn}, and specifically from the scaling behaviour of the fish diagram, for more details see~\Cref{app:rho4}. We remark that the sum rule~\labelcref{eq:SpecSum} enforces 
\begin{align} 
	\lim_{\lambda\to \infty}\rho(\lambda) < \frac{1}{\lambda^{2+\epsilon}}\,, 
\end{align}
for pole masses $m_\textrm{\tiny{pole}}>0$ and infinitesimal positive $\epsilon$. Hence, the sum rule is $k$-independent, and~\labelcref{eq:scalingrho} cannot hold true for asymptotically large spectral values.

\subsubsection{Spectral scaling}
\label{sec:SpectralScaling} 

\begin{figure*}
	\centering
	\begin{minipage}{0.99\linewidth}
		\centering
		\begin{subfigure}[t]{0.48\linewidth}
			\includegraphics[width=\textwidth]{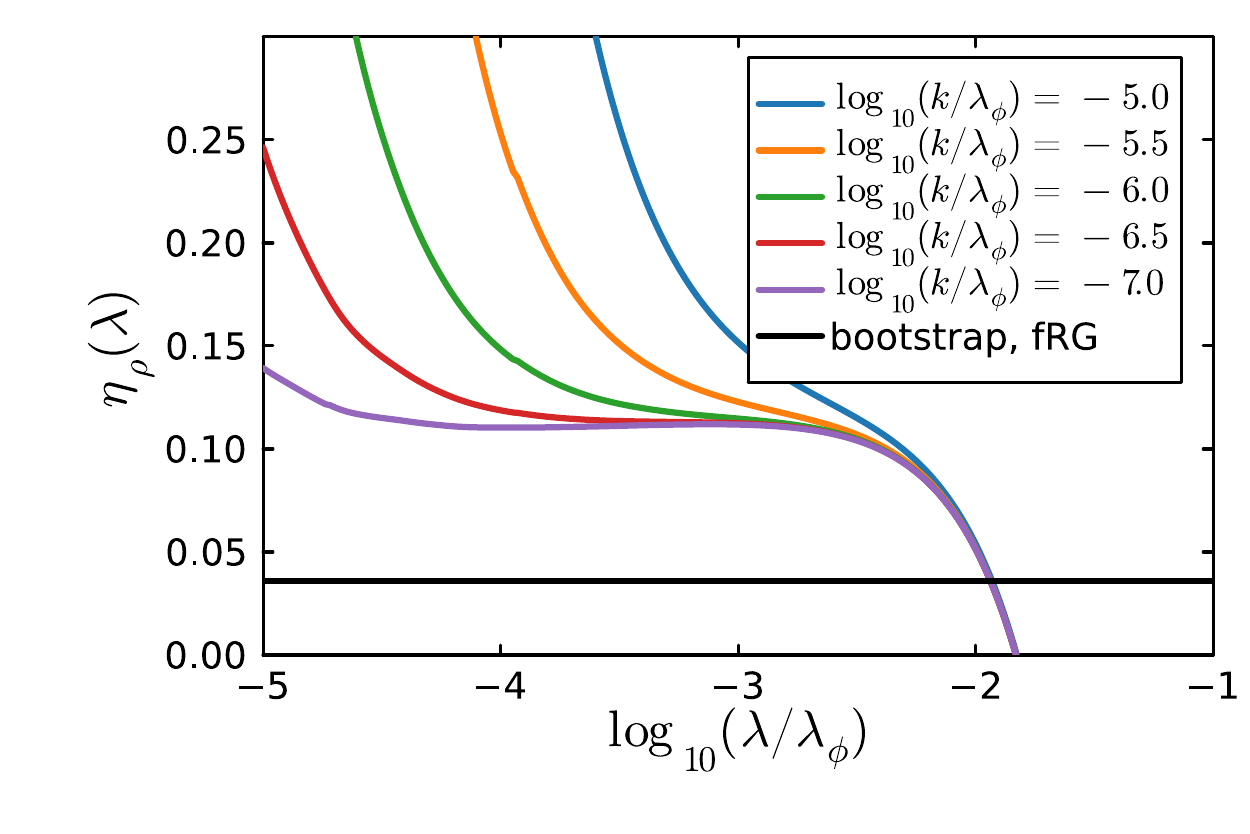}
			\caption{Scaling exponent $\eta_\rho(\lambda)$,~\labelcref{eq:etarholambda}, as a function of $\lambda$. \hspace*{\fill}}
			\label{fig:etarho}
		\end{subfigure}\hfill
		\begin{subfigure}[t]{0.48\linewidth}
			\includegraphics[width=\textwidth]{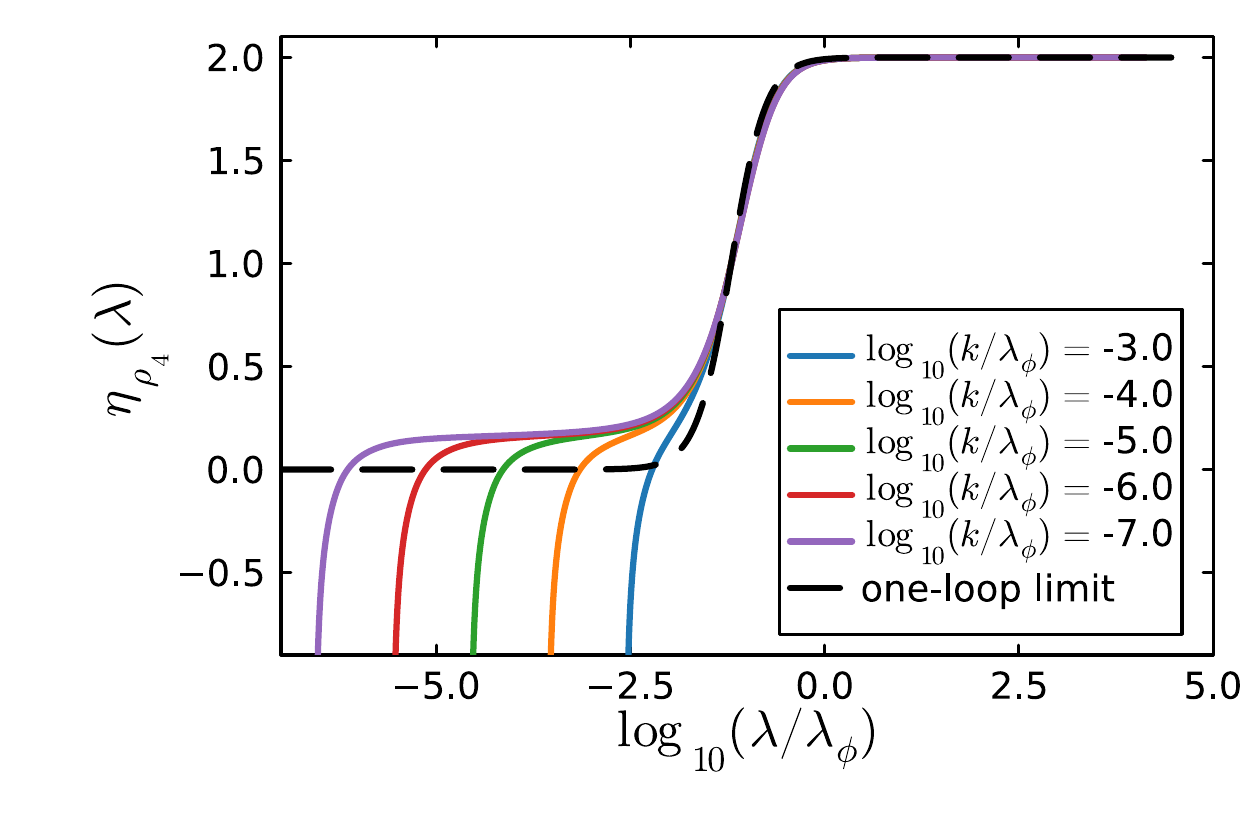}
			\caption{Scaling exponent $\eta_{\rho_4}(\lambda)$,~\labelcref{eq:etarho4lambda}, as a function of $\lambda$. As comparison, we show the one-loop limit as defined in~\labelcref{eq:OneLooprho4} \hspace*{\fill}}
			\label{fig:etarho4}
		\end{subfigure}
		\caption{Sliding scaling exponents $\eta_i(\lambda)$ with $i=\rho,\, \rho_4$ as functions of $\lambda/\lambda_\phi$. The scaling regime emerges for $\lambda/\lambda_\phi \lesssim 10^{-2}$. We also depict the quantitative reference results by the thick black line: conformal bootstrap, $\eta=0.03631 (3)$, \cite{El-Showk:2014dwa}, and fRG, $\eta=0.0361 (3)$ from \cite{Balog:2019rrg}. 
			\hspace*{\fill} } 
		\label{fig:etalambda}
	\end{minipage}
\end{figure*}

We proceed with a discussion of the spectral functions $\rho(\lambda)$ and $\rho_4(\lambda)$ as well as the respective critical exponents defined by the methods (1,2), discussed at the beginning of~\Cref{sec:Results}. 
Our numerical results for the spectral functions are shown in the doubly logarithmic plots in~\Cref{fig:rho+rho4}. 
The emergence of an increasing scaling window in the spectral regime $\lambda/\lambda_\phi \lesssim 10^{-2}$ with power law decays~\labelcref{eq:Scalingrhos} for successively smaller pole masses is clearly visible. 
The respective critical exponents can be extracted from~\Cref{fig:etarho} and~\Cref{fig:etarho4}, where we show the sliding scaling exponents $\eta_i(\lambda)$ with $i=\rho,\rho_4$. These sliding scaling exponents are defined by logarithmic spectral derivatives of the logarithms $\log \rho,\log( -\rho_4)$. We also subtract the canonical scalings, to wit,
\begin{subequations}
\label{eq:etalambdas} 
\begin{align} 
	\eta_\rho(\lambda) :=  2+ \frac{\lambda}{\rho(\lambda)} \frac{\partial \rho }{\partial \lambda} \,, 
	\label{eq:etarholambda}
\end{align} 
and 
\begin{align} 
	\eta_{\rho_4}(\lambda) := 1 -  \frac{\lambda}{\rho_4(\lambda)} \frac{\partial \rho_4}{\partial \lambda} \,.
	\label{eq:etarho4lambda}
\end{align} 
\end{subequations}
For the scaling spectral functions~\labelcref{eq:Scalingrhos}, the sliding scaling exponents $\eta_i(\lambda)$ reduce to the critical exponent $\eta$ and $2\,\eta$, respectively. We define the critical exponent $\eta_i$ by the plateau value of $\eta_i(\lambda)$ in~\Cref{fig:etalambda} when taking the limits $k\to 0$ and then $\lambda\to 0$. Both plateaus extend successively into the infrared in the scaling limit with $\lambda_\phi /m_\phi \to \infty$.

The upper boundary of the plateaus stays constant: the $s$-channel resummation for the four-point function in the current approximation leads to a decaying momentum dependence of the flow of the two-point function. This is related to the UV behaviour of the four-point function, which approaches the classical coupling $\lambda_\phi$ at high momenta due to the decay of the loop term in~\labelcref{eq:G4Full}. This behaviour ensures the sum rule as already discussed below~\labelcref{eq:Scalingrhos}. Note also that the intrinsic scale, at which the four-point function loses scaling, depends on the initial condition, i.e., the initial classical coupling $\lambda_\phi$. For a detailed discussion  see~\Cref{app:rho4}. This freezing of the flow explains the position of the upper boundary. The elimination of the intrinsic scale requires the feedback of non-trivial momentum dependencies of the four-point function as well as a more sophisticated initial condition. For a respective discussion see~\Cref{app:fullpotential}.

In conclusion, we compute the critical exponents in the frequency regime in which the single-particle and $s$-channel vertex spectral functions approach a scaling form. The critical exponents $\eta_i$ are readily extracted from the plateau values of $\eta_i(p)$, see~\labelcref{eq:etalambdas} and~\Cref{fig:etalambda}, for $\lambda/\lambda_\phi\lesssim 10^{-3}$. These plateaus extend towards $\lambda\to 0$ for $m_\phi/\lambda_\phi\to 0$. To estimate the value of the anomalous dimension $\eta$ at the fixed point, we perform an extrapolation of $\eta_\rho$ and $\eta_{\rho_4}$ to $k=0$ and subsequently to $\lambda= 0$; see~\Cref{app:Extrapolation} for details. The results are given by
\begin{subequations}
\label{eq:etaNums}
\begin{align}
\eta_\rho = 0.101^{+0.004}_{-0.028}\,,
\label{eq:etaNum}
\end{align}
and
\begin{align}
\frac{\eta_{\rho_4}}{2} = 0.077^{+0.002}_{-0.003}\,.
\label{eq:eta4Num}
\end{align}
\end{subequations}
The two values agree within the error bars. The momentum dependence of the four-point function is entirely determined by the scattering tail of the spectral function of the propagator. Thus, this tail is the only possible source of scaling in $\rho_4$. As shown in~\Cref{fig:Zk}, the pole contribution still carries approximately half of the spectral weight at $k/\lambda_\phi \approx 10^{-6}$, and therefore contributes significantly to $\rho_4$. Note that the flow induced solely by the pole of the propagator does not contribute to $\eta_{\rho_4}$, as can be seen from the black dashed line in~\Cref{fig:etarho4}. This analysis suggests that the result~\labelcref{eq:eta4Num} is lower than the actual value for $\eta$. The critical exponent $\eta_{\rho_4}$ of the four-point spectral function should therefore be interpreted primarily as a consistency check and a lower bound for the critical exponent $\eta_\rho$ of the propagator. In turn, the latter provides a more accurate estimate of the anomalous dimension of the theory.

The agreement between these two values further validates the computation of scaling spectral functions using the \textit{spectral} Callan–Symanzik flow in $\varphi^4$ theory. Our results are consistent with those obtained using spectral Dyson–Schwinger equations in a comparable truncation in the broken phase ($\eta \approx 0.11$ in~\cite{Eichmann:2023tjk}), and with functional renormalisation group (fRG) results on the Keldysh contour ($\eta = 0.0988$ in~\cite{Roth:2023wbp}).

We conclude the first part of this section with a brief discussion of the results in~\labelcref{eq:etaNums} in light of the critical exponent $\eta \approx 0.036$ for the three-dimensional Ising model. Reference values include the conformal bootstrap result $\eta = 0.03631(3)$~\cite{El-Showk:2014dwa}, and the fRG result $\eta = 0.0361(3)$~\cite{Balog:2019rrg}. A significant part of the deviation from these benchmark values can be attributed to the $\phi^4$ approximation of the effective potential. This will be illustrated through a comparison of critical exponents in LPA${}^\prime$ in~\Cref{sec:LPAprime}. This deficiency is readily resolved by implementing a full effective potential along the lines of \cite{Helmboldt:2014iya}. For a respective discussion see~\Cref{app:fullpotential}.

\subsubsection{Cutoff scaling}
\label{sec:flowofZ}

\begin{figure}[t]
\centering
\begin{minipage}{0.99\linewidth}
\includegraphics[width=\textwidth]{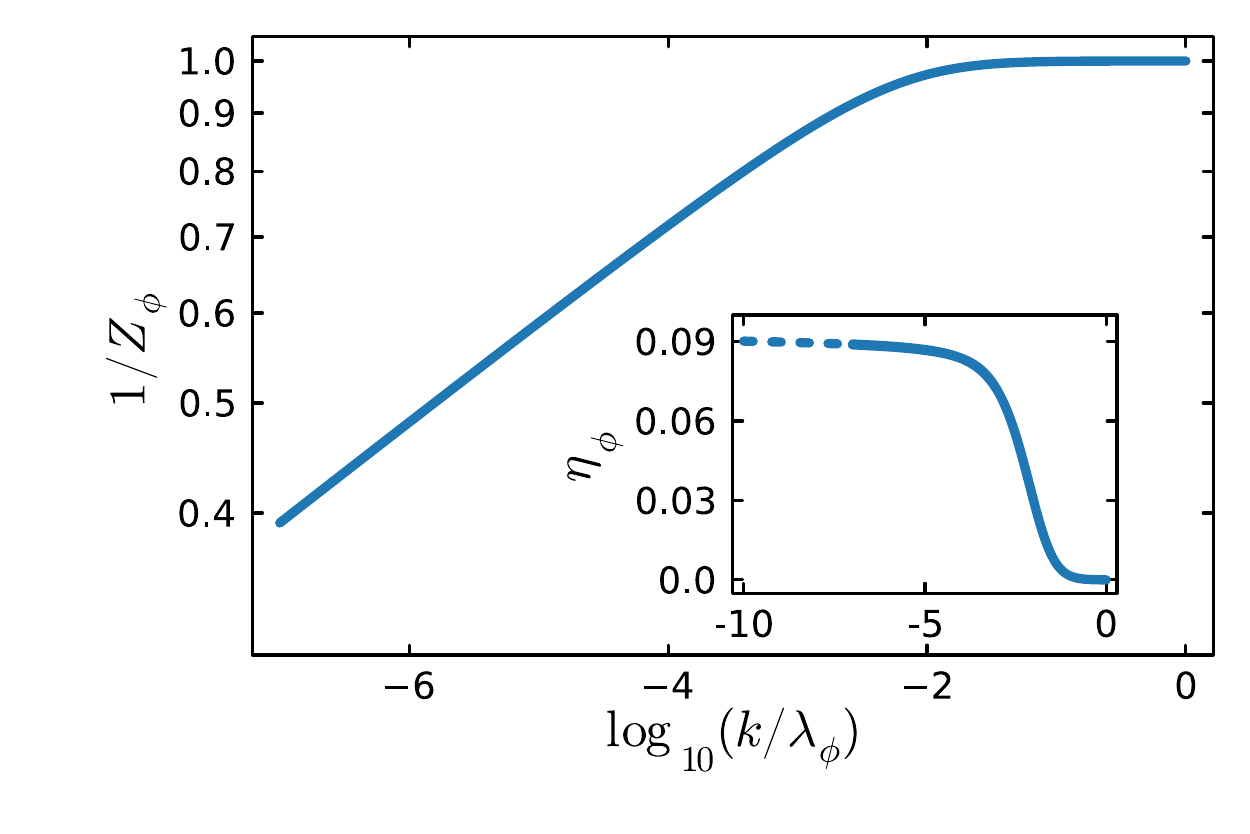}
\caption{Flow of the spectral weight $1/Z_\phi$ of the pole contribution as a function of the pole mass $m_{\text{\tiny{pole}}}=k$. For large pole masses with $k/\lambda_\phi\to \infty$, $Z_{\phi}$ approaches unity, which reflects its UV-irrelevance. For $k/\lambda_\phi\to 0$, the tail successively carries more of the spectral weight, and the pole contribution vanishes with a power law. The inset shows the anomalous dimension $\eta_{\phi}$,~\labelcref{eq:etaphik}, as a function of the pole mass. For $k\to 0$ it approaches the critical exponent $\eta$. \hspace*{\fill}}
\label{fig:Zk}
\end{minipage}
\end{figure}

With~\labelcref{eq:etaNum,eq:eta4Num} we have obtained $\eta$ with the methods (1,2), discussed at the beginning of~\Cref{sec:Results}. The third method consists of using the $k$-scaling of the wave function at the pole as a proxy for the momentum and spectral scaling. The result for ${Z_{\phi,k}=Z_{\phi,k}(p^2=-k^2)}$ as a function of the pole mass $m_{\text{\tiny{pole}}}=k$ is depicted in~\Cref{fig:Zk}. For large pole masses, $k\to \infty$, the wave function $Z_\phi$ approaches unity: it is a UV-irrelevant coupling, and accordingly its flow dies out. Moreover, the theory approaches the classical one, and the full spectral weight is carried by the pole, while the scattering tail vanishes throughout. This corresponds to the perturbative limit, see~\Cref{app:initial conditions}.

In the limit of vanishing pole masses, $k/\lambda_\phi\to 0$, the theory approaches the scaling regime around the phase transition.
Signals for scaling can be found for $k/\lambda_\phi \lesssim 10^{-2}$, where the wave function enters a power-law regime. This is rather similar to the scaling regime in the spectral functions $\rho,\ \rho_4$ with $\lambda/\lambda_\phi \lesssim 10^{-2}$, see~\Cref{fig:rho,fig:rho4}. For $k/\lambda_\phi\to 0$, the theory is increasingly dominated by scattering processes, and the spectral weight $1/Z_\phi$ of the pole contribution is successively suppressed and vanishes for $k=0$. As the spectral weight of the scattering tail is increasingly dominated by the scaling part, the wave function $Z_\phi$ at the pole has to scale with $k^{-\eta}$. Similar to~\labelcref{eq:etarholambda} and~\labelcref{eq:etarho4lambda}, we define
\begin{align}
\eta_{\phi,k} = -\frac{\partial_t Z_{\phi,k}}{Z_{\phi,k}}\,.
\label{eq:etaphik}
\end{align}
We remark that $\eta_{\phi,k}$ differs from the standard anomalous dimension used in fRG studies: the latter is typically defined as the total $t$-derivative at a fixed momentum,
\begin{align}
\eta_{\phi}(p)= - \frac{\partial_t Z_{\phi}(p)}{Z_{\phi}(p)}\,,
\label{eq:etaphip}
\end{align}
often evaluated at $p=0$. In turn,~\labelcref{eq:etaphik} is given by ${\eta_{\phi} =\eta_{\phi}(p) - p\partial_p Z_{\phi}(p)}$, evaluated at $p^2=-k^2$.

The anomalous dimension $\eta_{\phi}$ is depicted in the inset of~\Cref{fig:Zk} as a function of the pole mass $m_{\text{\tiny{pole}}}=k$. \Cref{fig:Zk} also shows $Z_{\phi}$, and the scaling limit is visible in both quantities. We read off the critical exponent $\eta$ as the limit $k\to 0$, with
\begin{align}
\eta=\lim_{k\to 0} \eta_{\phi} = 0.095(9)\,.
\label{eq:etaetaphik}
\end{align}
\Cref{eq:etaetaphik} agrees well with the critical exponent obtained from the single-particle spectral function,~\labelcref{eq:etaNum}. In order to appreciate its accuracy, it should be contrasted with similar approximations in the derivative expansions, which is done in the next section.

\subsection{Benchmarks and extensions}
\label{sec:LPAprime}

We close this Section with a discussion of the systematics of the present approximation, its embedding into the landscape of existing results, and of systematic improvements: \\[-2ex] 

In the present work we aimed at the computation of fully momentum-dependent real-time correlation functions with an emphasis on the approach to the scaling regime around a second order phase transition. 
The respective critical exponent $\eta$ is one of the outcomes of this analysis, but it was not the primary target. Still, we will embed the present results and the underlying approximation in the landscape of dedicated fRG computations of critical exponents. 
Typically, these computations are done within a fixed point analysis of the fRG which allows for the most direct access to critical physics but does not cover the interface to the non-universal regime. 
These results have mostly been achieved within the derivative expansion, for an overview see the review \cite{Dupuis:2020fhh}. 
In terms of comparability, the current approximation with momentum-dependent but field-independent dressings fits to the LPA${}^\prime$ approximation (effective potential and cutoff dependent wave function) that lies in between the 0th  order of the derivative expansion (with an effective potential (LPA)), and the first order (effective potential and field-dependent wave function).  

In contradistinction to LPA${}^\prime$ we have included fully momentum-dependent propagators and and $s$-channel vertices, but only considered the effective potential up to the order $\phi^4$. Moreover, we have approached the scaling regime from the symmetric regime. 
However, the fixed point analysis reveals a finite dimensionless expectation value of the field with the fixed point value $\bar\rho^*\neq 0$ and $\bar\rho=Z_\phi \phi^2/(2 k)$, see~\labelcref{eq:Dimlessurho} in~\Cref{app:LPAbenchmark}. This suggests that scaling properties converge faster in the broken regime. 
Furthermore, we have already mentioned before that a comparison between LPA${}^\prime$ and the present approximation scheme, or rather the specific order of this scheme used here, has been done for a Yukawa model in \cite{Helmboldt:2014iya}, concentrating on non-universal physics. There it was found, that
pole masses and further observables agreed very well on the percent level, if fully momentum-dependent computations and LPA${}^\prime$ computations with full effective potentials are compared. 
We expect that this agreement worsens in the scaling regime with its algebraic (non-local) momentum decays. 
In any case, LPA${}^\prime$ can be embedded in the present approximation as a lower order approximation. 

Finally, the present computation uses the on-shell renormalisation scheme instead of the standard one: With on-shell renormalisation all quantities are measured directly in the physical correlation length $\xi \propto 1/k$ with the pole mass $m_\textrm{pole} =k$ in contradistinction to $\xi\propto k^{-\eta}$ in the standard fRG renormalisation scheme. This scheme, the MOM${}^2$-scheme has similarities to a MOM-scheme commonly used in perturbation theory and Dyson-Schwinger equations, but also carries some differences, see \cite{Gao:2021wun}. In any case we expect different convergence pattern for both schemes.

Bearing these differences and similarities in mind, we compare the present results to LPA${}^\prime$ results of a fixed point analysis within a Taylor expansion of the full effective potential. The respective computations in LPA${}^\prime$ within the standard RG scheme for different regulators can be found in the literature, see the review \cite{Dupuis:2020fhh}. We start the discussion with the comparison of the present truncation with $V_\textrm{eff}^{(n>2)}\equiv0$ to a standard Euclidean fixed point analysis in LPA${}^\prime$ with standard and on-shell renormalisation in the same order of the Taylor expansion. The computation is deferred to~\Cref{app:LPAbenchmark}. We are led to the critical exponent
\begin{align} 
 \eta_{\textrm{LPA}^\prime} = 0.1600\,.
\label{eq:etaLPAprime}
\end{align}
\Cref{eq:etaLPAprime} has to be compared with $\eta\approx 0.1$ from~\labelcref{eq:etaNums} and~\labelcref{eq:etaetaphik}, see~\Cref{tab:etaComparison}. 
Evidently, the inclusion of full momentum or spectral dependences yields significantly better results, even though the fixed point analysis is bound to have the better  convergence due to expanding about the fixed point value of the field with $\bar\rho^*>0$. 
The inclusion of higher order scatterings in the effective potential leads to a quick convergence for the critical exponent of the fixed point analysis in LPA${}^\prime$ with the CS-regulator:   $\eta = 0.0802$, see~\Cref{app:LPAbenchmark} for details.
\begin{table}[t]
	\centering
	\begin{tabular}{| >{\centering\arraybackslash} c| >{\centering\arraybackslash} c| >{\centering\arraybackslash} c|}
		\hline
		&  Correlation function & $\eta$  \rule[-1.3ex]{0pt}{4ex}\\
		\hline\hline
		& propagator & $0.101^{+0.004}_{-0.028}$ \rule[-1.3ex]{0pt}{4ex}\\
		\cline{2-3}
		This work & four-point function & $0.077^{+0.002}_{-0.003}$   \rule[-1.3ex]{0pt}{4ex}\\
		\cline{2-3}
		& flow of $Z_\phi$ & 0.095(9)\rule[-1.3ex]{0pt}{4ex}\\
		\hline 
		LPA${}^\prime$ Benchmarks & $V_\textrm{eff}^{(n>2)}=0$  & $0.16$ \rule[-1.3ex]{0pt}{4ex}\\
		\cline{2-3}
		         & $V_\textrm{eff}^{(n>9)}=0$ & $0.0802$ \rule[-1.3ex]{0pt}{4ex}\\
		\hline
	\end{tabular}
	\caption{Critical exponent $\eta$ from different correlation functions in the present work. The four-point function result should be interpreted as a cross-check to the propagator one. The deviation of less than $1\sigma$ renders it fulfilled. We also list benchmarks from LPA${}^\prime$ with the CS-regulator. Here, $V_\textrm{eff}^{(n)}=\partial_\rho^n V_\textrm{eff}$, see~\labelcref{eq:Veff} and below. The LPA${}^\prime$ results are obtained within the fixed point analysis, see also~\Cref{app:LPAbenchmark} where we also studied the convergence of the Taylor expansion in~\Cref{tab:LPAResults}. \hspace*{\fill}}
	\label{tab:etaComparison}
\end{table}

The large improvement upon including the momentum dependence is expected for the CS-regulator from functional optimisation theory, see \cite{Litim:2000ci, Litim:2001dt, Litim:2002cf, Pawlowski:2005xe, Pawlowski:2015mlf}: the CS-regulator is not optimised in terms of a convergence of an expansion in momentum dependences as it collects contributions in the full (loop) momentum regime. Accordingly, we expect better approximations of the full momentum dependence to lead to sizeable improvements. In turn, regulators that are optimised for approximation schemes relying on expansions in momentum dependences, such as the derivative expansion, lead to a more rapid convergence. This is evident within a comparison to LPA and LPA${}^\prime$ results with the flat or Litim regulator, \cite{Litim:2000ci, Litim:2001dt, Litim:2002cf}, which is the optimised one for LPA but not beyond~\cite{Pawlowski:2005xe}. With the same approximation to the effective potential as used here, we obtain 
\begin{align} 
	\eta_\textrm{flat} = 0.0546\,, 
	\label{eq:etaLPALitim4}
\end{align}
and with a full effective potential, we obtain $\eta=0.0443$. 

The results with the flat regulator point at a huge quantitative advantage of the optimal choice of the regulator in low orders of the derivative expansion. However, the CS-regulator allows for the use of the on-shell renormalisation scheme, whose implementation for momentum-dependent regulators is intricate. Moreover, this intricacy is even more challenging for non-analytic regulators such as the flat one. It is very suggestive that the on-shell renormalisation scheme~\labelcref{eq:onshellrenSymmetric} with $m_{\text{\tiny{pole}}}=k$ leads to a far better convergence of observables and undoes at least part of the momentum non-localities. This will be discussed elsewhere.

\section{Conclusion}
\label{sec:Conclusions}

We have investigated the real-time scalar $\phi^4$ theory in 2+1 dimensions in and outside the scaling regime close to the phase transition. Our method of choice was the spectral Callan-Symanzik equation, and the present setup can be readily used within the mesonic sector of full functional QCD within the fRG approach. In particular, we have computed the single-particle spectral functions and that of the four-point function in the $s$-channel, see~\Cref{fig:rho+rho4} in~\Cref{sec:Results} and~\Cref{fig:rho4} in~\Cref{app:rho4}. The results indicate that the scaling regime is reached for ratios of the pole mass $m_{\text{\tiny{pole}}}$ and the classical coupling  $\lambda_\phi$ with $m_{\text{\tiny{pole}}}/\lambda_\phi \lesssim 10^{-2}$, which signals a very small scaling regime. This result agrees well with the respective findings in Euclidean computations. 

In the scaling regime, these results also allow us to compute the critical exponent $\eta$ of the Ising universality class, see~\Cref{tab:etaComparison}. We have discussed in~\Cref{sec:LPAprime}, how these results compare to further ones, mostly obtained within the derivative expansion but also within approximations related to the one used here. 

Further improvements are work in progress, ranging from the inclusion of the full effective potential to the extension to the quark-meson sector of QCD with physics-informed flows, \cite{Ihssen:2023nqd, Ihssen:2024ihp}. We hope to report on respective results soon. \\[-4ex]

\subsection*{Acknowledgements}

We thank Gernot Eichmann, Andrés Gómez, Jan Horak and Nicolas Wink for discussions and collaborations on related projects. This work is done within the fQCD-collaboration~\cite{fQCD}, and we thank the members for discussion and collaborations on related projects. This work is funded by the Deutsche Forschungsgemeinschaft (DFG, German Research Foundation) under Germany's Excellence Strategy EXC 2181/1 - 390900948 (the Heidelberg STRUCTURES Excellence Cluster) and under the Collaborative Research Centre SFB 1225 (ISOQUANT). 
\appendix

\begingroup
\allowdisplaybreaks

\begin{figure*}[ht]
	\centering
	\begin{minipage}{0.99\linewidth}
		\centering
		\begin{subfigure}[t]{0.48\linewidth}
			\includegraphics[width=\textwidth]{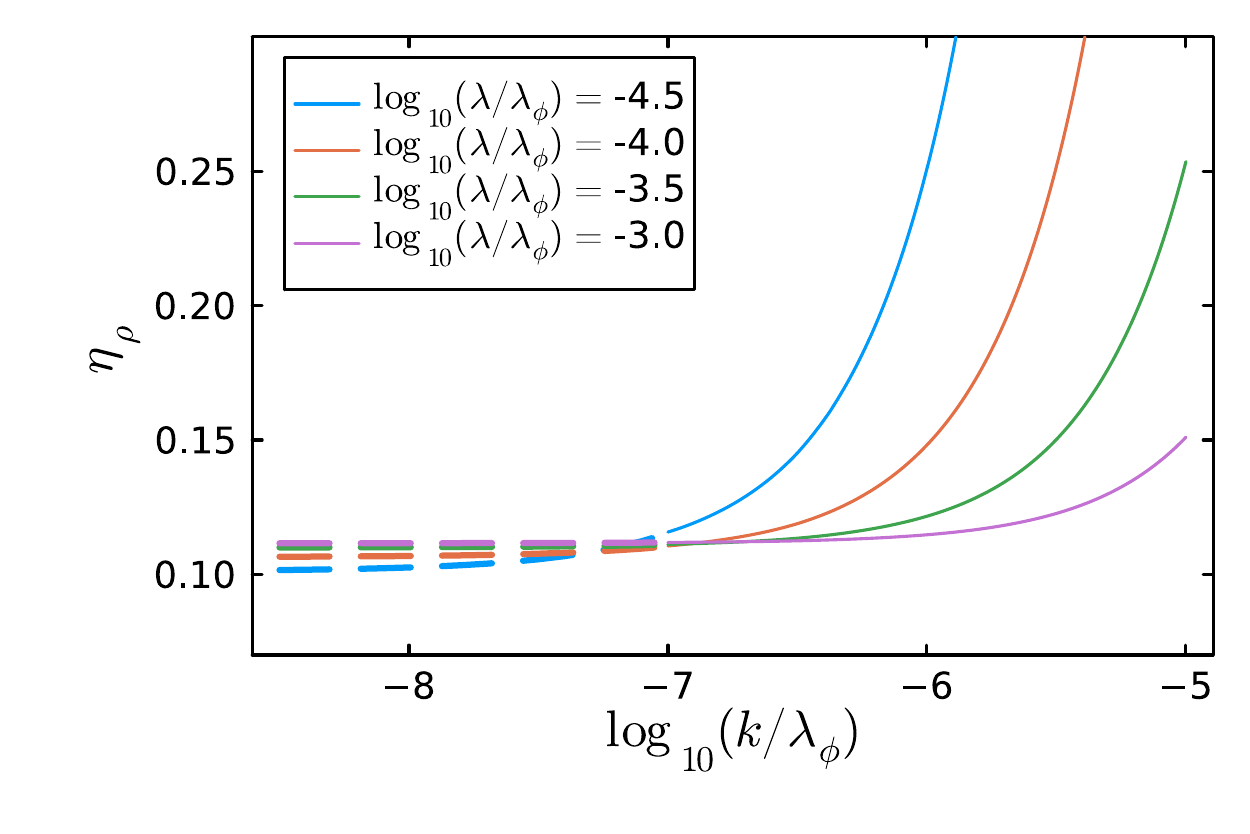}
			\caption{Scaling exponent of the propagator spectral function. \hspace*{\fill}}
			\label{fig:rhoExtrapolation}
		\end{subfigure}\hfill
		\begin{subfigure}[t]{0.48\linewidth}
			\includegraphics[width=\textwidth]{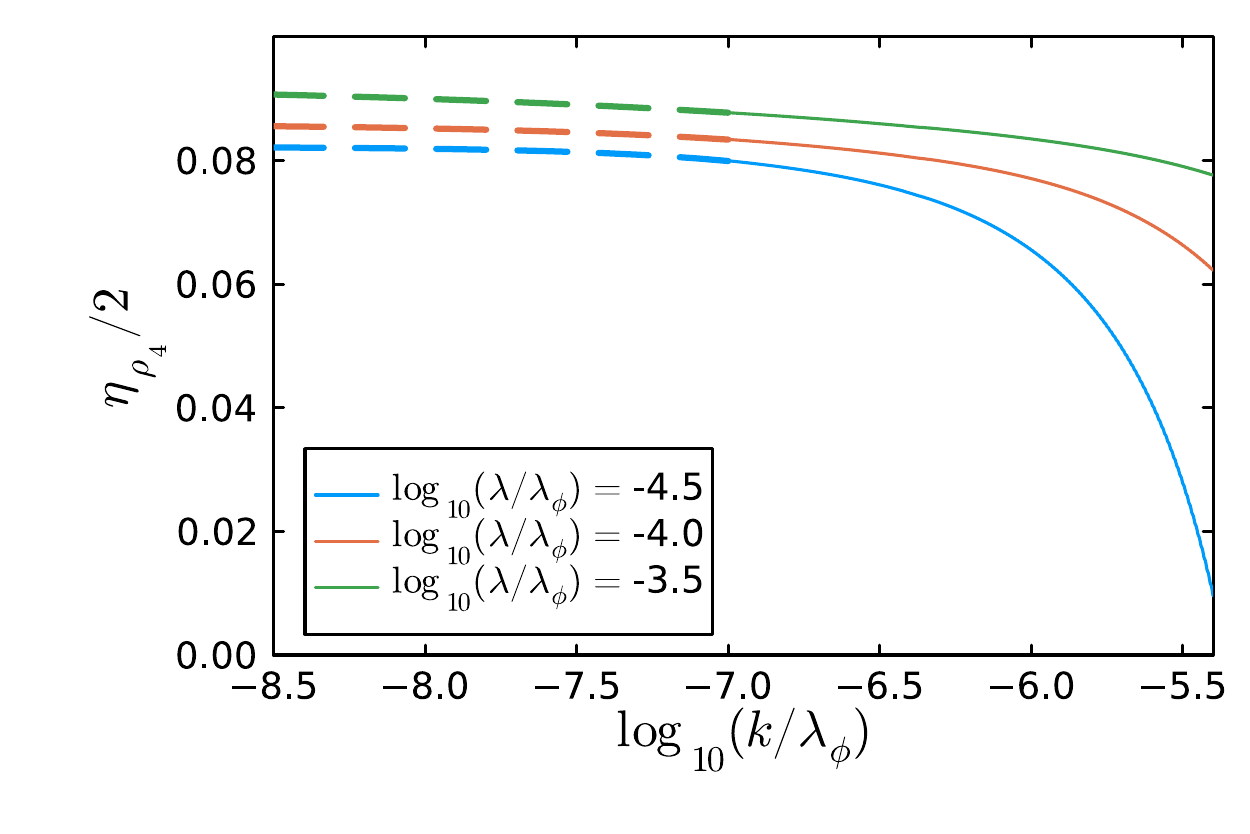}
			\caption{Scaling exponent of the four-point spectral function.  \hspace*{\fill}}
			\label{fig:rho4Extrapolation}
		\end{subfigure}
		\caption{Logarithmic derivatives $\eta_\rho$ and $\eta_{\rho_4}$ at different spectral parameters $\lambda$ (see~\labelcref{eq:etarholambda} and~\labelcref{eq:etarho4lambda}) as a function of the pole mass $m_\text{\tiny{pole}}=k$. We extrapolate towards $k=0$. $\eta_\rho$ is extrapolated with a parabola and $\eta_{\rho_4}$ with a fifth-order polynomial. For details, see~\Cref{app:ExtrapolationK}. The limit $k\to 0$ is shown in~\Cref{fig:ExtrapolationLimit}. 
			\hspace*{\fill} } 
		\label{fig:Extrapolation}
	\end{minipage}
\end{figure*}
%
\section{Anomalous dimension $\eta_{\phi}$}
\label{app:FlowConstants} 

In this Appendix, we provide the relation used for the determination of the anomalous dimension $\eta_\phi$ defined in~\labelcref{eq:etaphimu}. We start with the parametrisation of $\Gamma^{(2)}(p)$ that follows from~\labelcref{eq:Gp}, 
\begin{align}  
	\Gamma^{(2)}(p) = Z_\phi(p)\left( p^2 +m_\text{\tiny{pole}}^2\right)\,, 
	\label{eq:Gamma2p} 
\end{align} 
where we use Euclidean momenta $p^2$. Throughout the derivation, we use the on-shell renormalisation condition~\labelcref{eq:onshellrenSymmetric}, leading to $m_\textrm{pole}^2=k^2$. The on-shell wave function is given by $Z_\phi=Z_\phi(p^2 = -k^2)$, which can be computed from the $p^2$-derivative of $\Gamma^{(2)}(p)$, 
\begin{align} 
	Z_\phi=\left. \frac{\partial \Gamma^{(2)}(p)}{\partial {p^2}}\right|_{p^2=-k^2}\,.
	\label{eq:DefofZphi}
\end{align}
The $t$-derivative of~\labelcref{eq:DefofZphi} hits both the explicit $k$-dependence of $\Gamma^{(2)}$ and the $k$-dependence of the momentum argument. This leads us to an explicit expression for $\eta_\phi$, 
\begin{align}
	\eta_\phi 	=\frac{1}{Z_\phi}\Biggl[\frac{\partial}{\partial {p^2}} \partial_t \Gamma^{(2)}(p)-2 k^2 \frac{\partial^2 \Gamma^{(2)}(p)}{(\partial p^2)^2}\Biggr]_{p^2=-k^2}\hspace{-.2cm}\,.
\label{eq:etaphiApp}
\end{align}
The second term in~\labelcref{eq:etaphiApp} can be expressed in terms of the spectral representation~\labelcref{eq:specrepprop} of the propagator~\labelcref{eq:Gp}. To that end, we use that 
\begin{align}
	\partial_{p^2}^2\Gamma^{2}(p)=(p^2+k^2)\partial_{p^2}^2Z_\phi(p) + 2\,\partial_{p^2}Z_\phi(p)\,.
	\label{eq:p2-derGamma2}
\end{align}
The first term in \labelcref{eq:p2-derGamma2} vanishes on-shell with $p^2= -k^2$. For the second term, we use that
\begin{align} \nonumber
\frac{\partial Z_\phi(p)}{\partial p^2} =&\, -Z_\phi^2(p)\frac{\partial}{\partial p^2}\Biggl[\frac{1}{Z_\phi(p)}\Biggr]\\[1ex]
=&\,- Z^2_\phi(p) \frac{\partial}{\partial p^2} \Bigl[ ( p^2 +k^2) \,G(p)\Bigr]\,.
\label{eq:Zphiprime1}
\end{align}
With the spectral representation~\labelcref{eq:specrepprop} of the propagator and~\labelcref{eq:rhoparam} we split the propagator in its pole and tail contributions,  
\begin{align}
	\frac{\partial Z_\phi(p)}{\partial p^2} =- Z^2_\phi(p) \frac{\partial}{\partial p^2} \Biggl[\frac{1}{Z_\phi} + ( p^2 +k^2) \,\int\limits_\lambda\frac{\tilde{\rho}(\lambda)}{\lambda^2+p^2}\Biggr] .
	\label{eq:Zphiprime2}
\end{align}
The on-shell residue $1/Z_\phi$ is momentum-independent. Hence, the one-particle pole of the propagator does not contribute to $\partial_{p^2}Z(p)$. Going on-shell, we arrive at 
\begin{align}
\left.\frac{\partial Z_\phi(p)}{\partial p^2}\right|_{p^2=-k^2} = -Z_\phi^2\int_\lambda\frac{\tilde{\rho}(\lambda)}{\lambda^2-k^2}\,.
\label{eq:Zphiprime3}
\end{align}
Note that the pole of the integrand lies outside the support of $\tilde{\rho}(\lambda)$. 
Putting these results together, we get 
\begin{align}
	\eta_\phi 	=\left.\frac{1}{Z_\phi}\frac{\partial}{\partial {p^2}} \partial_t \Gamma^{(2)}(p)\right|_{p^2=-k^2}\hspace{-.2cm} +4 k^2 Z_\phi\int_\lambda\frac{\tilde{\rho}(\lambda)}{\lambda^2-k^2}\,.
	\label{eq:etaphiAppFull}
\end{align}
The first term in~\labelcref{eq:etaphiAppFull} is readily derived from the spectral form of $\partial_t \Gamma^{(2)}(p)$  in~\labelcref{eq:FlowG2Final} with an analytic $p^2$-derivative. However,~\labelcref{eq:FlowG2Final} also depends on $\eta_\phi$, and a respective resummation leads us to the final relation, 
\begin{align}
	\eta_\phi=\left. \frac{-k^2\frac{\partial D_{\text{dyn}}^{\text{tad}}(\omega)}{\partial {\omega^2}}+4k^2Z_\phi^2\int_\lambda\frac{\tilde{\rho}(\lambda)}{\lambda^2-m_{\text{\tiny{pole}}}^2}}{1-\frac{k^2}{2}\frac{\partial D_{\text{dyn}}^{\text{tad}}(\omega)}{\partial {\omega^2}}}\right|_{\omega^2=m_{\text{\tiny{pole}}}^2}\hspace{-.5cm}\, .
\end{align}
Note that the frequency derivatives can be taken analytically. For the analytic expressions of the diagrams  see~\Cref{app:analyticintegrands}. This avoids any instabilities introduced by numerical derivatives.

\section{Scaling of $\rho_4(\lambda)$}
\label{app:rho4}

In this Appendix, we discuss the scaling limits of the $s$-channel spectral function of the four-point scattering vertex. The respective numerical results are shown in~\Cref{fig:rho4} and~\Cref{fig:etarho4}. The first possible scattering is $2\to 2$ scattering, leading to an onset at $m_{\text{\tiny{scat}}}=2m_{\text{\tiny{pole}}}$, as is visible in~\Cref{fig:rho4}. For small momenta or spectral values, both $\Gamma^{(4)}$ and $\rho_4$ approach a power law, described by~\labelcref{eq:scalingrho4}. Notably, both the prefactor and the exponent of these power laws are independent of $\lambda_\phi$. For very large momenta, the four-point function approaches its classical limit $\Gamma^{(4)}(\omega\to\infty)=\lambda_\phi$. In turn, the spectral function decays with $\omega^{-1}$, see~\Cref{fig:rho4}. This decay is produced by the sub-leading one-loop behaviour of $\Gamma^{(4)}$ for large momenta. The large momentum behaviour entails that $\rho_4(\lambda)$ approaches the respective one-loop spectral function for $\lambda\to\infty$: For illustration we compute the one-loop result with classical spectral functions in $D_{\textrm{\tiny{fish}}}$ in~\labelcref{eq:Dfish} with a vanishing mass, $\rho(\lambda)= 2 \pi\delta(\lambda^2)$ in~\labelcref{eq:rhoparam}. Then, $\Gamma^{(4)}$ reduces to 
\begin{align}
	\Gamma^{(4)}(\omega)=\frac{\lambda_\phi}{1+\imag\frac{\lambda_\phi}{16\omega}}\,,
\end{align}
and the $s$-channel spectral function $\rho_4$ reads  
\begin{align}
	\rho_4(\omega) =\frac{\lambda_\phi}{\frac{16\omega}{\lambda_\phi}+\frac{\lambda_\phi}{16\omega}}\, .
	\label{eq:OneLooprho4}
\end{align}
In the scaling limit, the propagator has the form $G(p)\propto(p^2)^{-(1-\eta/2)}$,~\labelcref{eq:scalingrho}. Inserting~\labelcref{eq:scalingrho} into the fish diagram $D_{\textrm{\tiny{fish}}}$ in~\labelcref{eq:Dfish}, leads us to 
\begin{align}
	D_{\textrm{\tiny{fish}}}(\omega)\propto (\omega^2)^{-1/2+\eta}\, , 
\end{align}
and the scaling law~\labelcref{eq:scalingrho4} for $\rho_4$ with the exponent $ 1- 2 \eta$. 

In~\Cref{fig:etarho4} we show $\eta_{\rho_4}(\lambda)$ for different pole masses $m_\textrm{pole}=k$ together with the one-loop result from~\labelcref{eq:OneLooprho4}, the back dashed line. Both, the scaling limit for small masses and $\lambda/\lambda_\phi \lesssim 10^{-2}$ as well as the one-loop limit for $\lambda/\lambda_\phi \gtrsim 10^{-2}$ are clearly visible. Moreover, the scaling limit grows for smaller pole masses towards $\lambda=0$. 

The convergence of the plateau towards the scaling exponent $\eta_{\rho_{4}}$ is discussed in~\Cref{app:Extrapolation}. In short, $\eta_{\rho_4}(\lambda,k)$ converges towards a value $\eta$ for $k\to 0$ and $\lambda\to 0$, but at $k/\lambda_\phi=10^{-7}$, this limit is not reached yet. We extract the limit $\eta$ from an extrapolation towards $k=0$ and subsequently $\lambda\to 0$. The respective uncertainty informs our error estimate.

\section{Comment on the critical exponent $\nu$}
\label{sec:nu}

Equilibrium second order phase transitions are governed by two independent critical exponents. A common choice for one of them in RG studies is $\eta$. It governs the scaling on the fixed point and is also computed in the present work. The other independent critical exponent governs the approach to the fixed point, and in fRG studies commonly $\nu$ is taken, the scaling of the correlation length,  
\begin{align}
	\xi\propto m_{\text{\tiny{pole}}}^{-1} \propto \tau^{-\nu}\, .
\end{align}
Here $\tau$ is the external tuning parameter, for example the temperature $T$, the magnetic field $B$ or simply the classical mass parameter of the theory. In the fRG approach, a further tuning parameter is introduced, the infrared cutoff scale $k$. However, on-shell renormalisation identifies the pole mass with the cutoff scale (for all regulator functions) and hence we have $\xi \propto 1/k$. This is part of the optimisation of an on-shell expansion scheme, see the discussion in~\Cref{sec:LPAprime}. Hence, it should be regarded as a feature not a bug. Still, it prevents the simple access to this critical exponent via the $k$-scaling. It can be extracted directly from the scaling with standard tuning parameters such as $T,B$. Moreover, it is also hidden in the flow of the counter term: the counter-term flow implements the flow of the on-shell renormalisation condition and hence the map from the standard tuning parameter $k$ to the trivial on-shell tuning parameter. We shall discuss the extraction of all critical exponents via physical tuning parameters, as well as clarifying of the on-shell tuning parameter and the standard one in a future work. 

\begin{figure*}[t]
	\centering
	\begin{minipage}{0.99\linewidth}
		\centering
		\begin{subfigure}[t]{0.48\linewidth}
			\includegraphics[width=\textwidth]{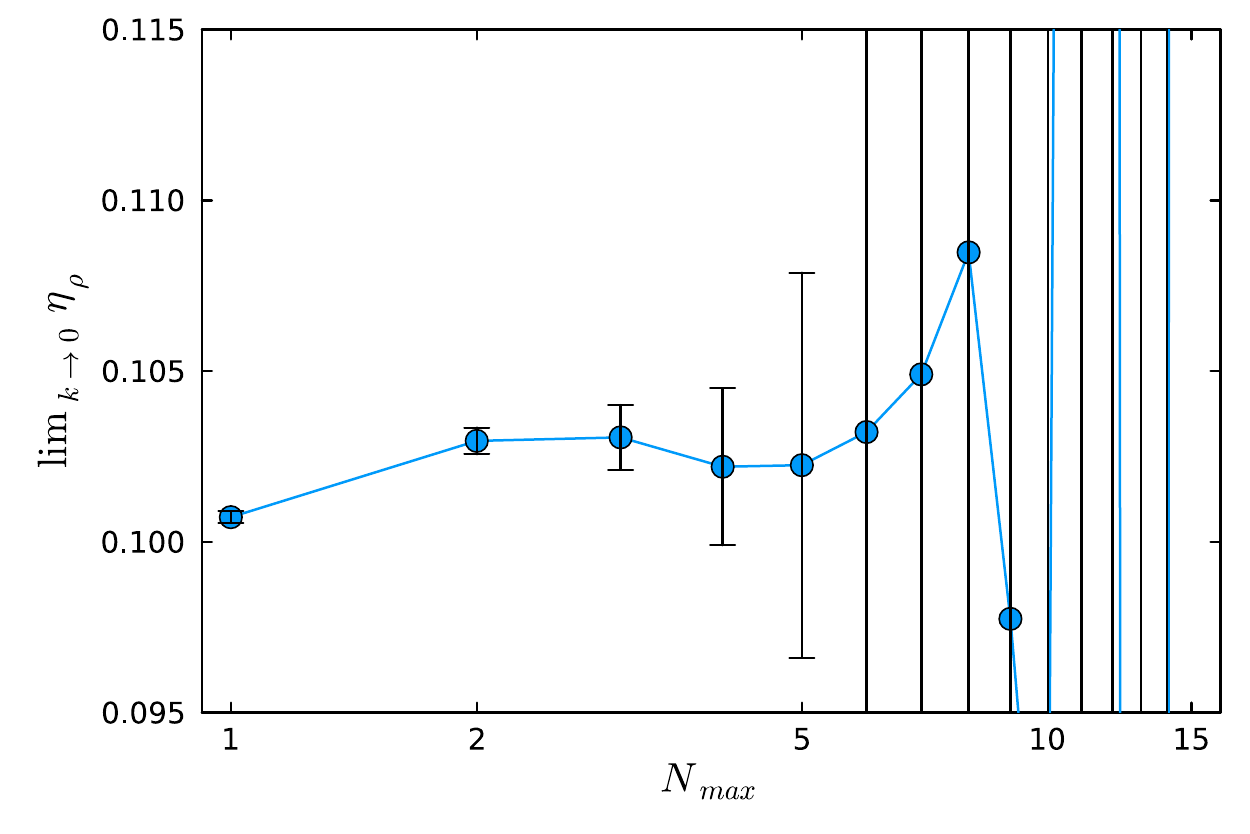}
			\caption{Limit $k\to 0$ of the Scaling exponent of the propagator spectral function with different polynomials used as extrapolation functions. \hspace*{\fill}}
			\label{fig:bestPolynomialFitLimit}
		\end{subfigure}\hfill
		\begin{subfigure}[t]{0.48\linewidth}
			\includegraphics[width=\textwidth]{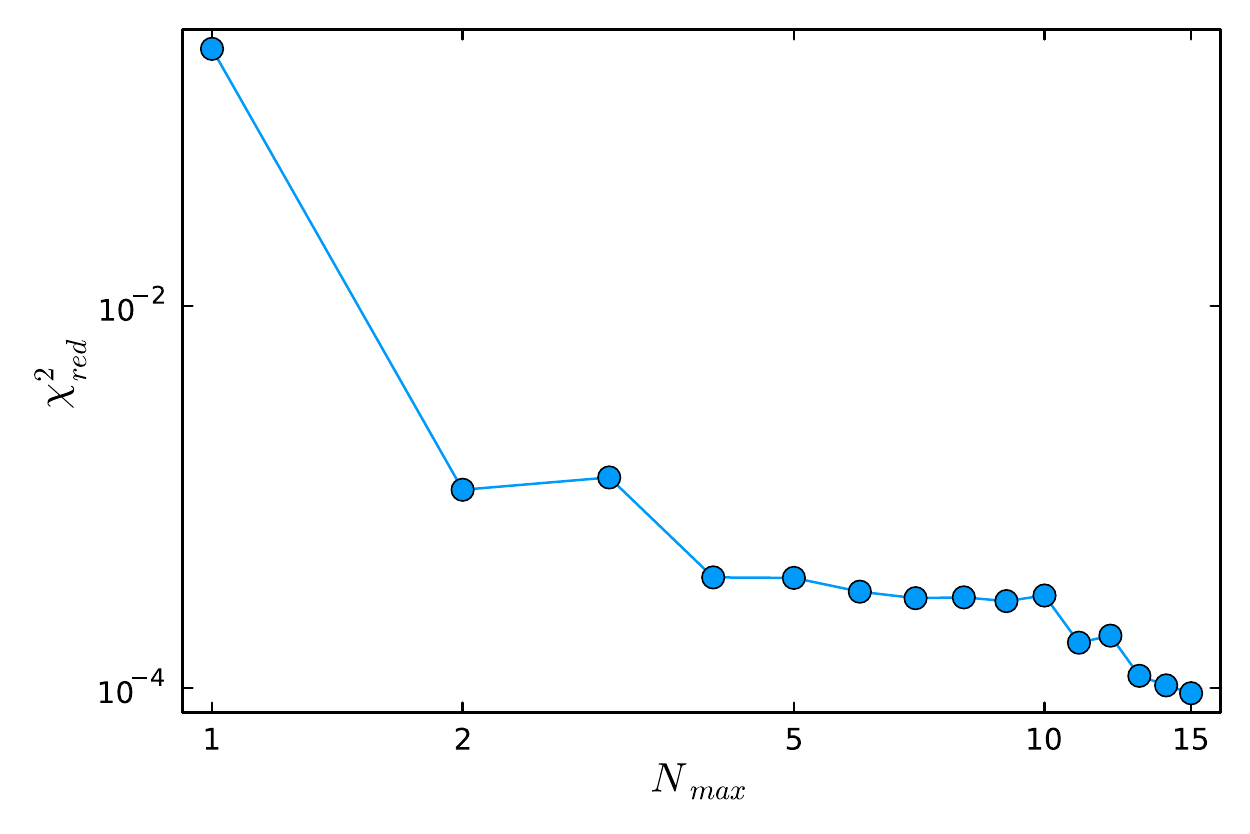}
			\caption{Reduced chi-square value $\chi^2_{\text{red}}$ for different polynomials used as extrapolation functions. \hspace*{\fill}}
			\label{fig:bestPolynomialFitChi2}
		\end{subfigure}
		\caption{Dependence of the extrapolation $k\to 0$ of the logarithmic derivative of the spectral function $\eta_\rho$ on the order $N_{\text{max}}$ of the polynomial used as fit function. We show the fits for $\log_{10}(\lambda/\lambda_\phi)=-4.5$, but the data sets for different $\lambda$ show a similar fitting behaviour. Very high order polynomials lead to highly fluctuating limits with large errors and very low $\chi^2_{\text{red}}$, indicating overfitting. The optimal fitting function is a polynomial with $N_{\text{max}}=2$.
			\hspace*{\fill} } 
		\label{fig:bestPolynomialFit}
	\end{minipage}
\end{figure*}
%

\section{Extrapolation of $\eta$ from spectral scaling}
\label{app:Extrapolation}

Extracting the value of the anomalous dimension $\eta$ from the spectral functions is not straightforward. While at the scale $k/\lambda_\phi=10^{-7}$ the plateaus in~\Cref{fig:etalambda} seem to be settled at a constant value, a more careful consideration of the height of said plateau as a function of the scale shows that this is not the case everywhere. \Cref{fig:Extrapolation} shows the evolution of the logarithmic derivatives $\eta_\rho$ and $\eta_{\rho_4}$ as defined in~\labelcref{eq:etarholambda} and~\labelcref{eq:etarho4lambda} at different points in the respective plateaus.

\subsection{Extrapolation towards $k=0$}
\label{app:ExtrapolationK}

As expected, at $k/\lambda_\phi=10^{-7}$ the value for the logarithmic derivative is not fully settled, especially at lower spectral parameters. To extract the limit $k\to 0$ we perform an extrapolation. We fit a polynomial of degree $N_{\text{max}}$ in $k$ to the logarithmic derivative at several spectral parameters $\lambda$ that lie in the respective scaling regions. For the fit region we choose 150 points between $\log_{10}(k/\lambda_\phi)=-6.25$ and $-7$.

\begin{figure*}[t]
	\centering
	\begin{minipage}{0.99\linewidth}
		\centering
		\begin{subfigure}[t]{0.48\linewidth}
			\centering
			\includegraphics[width=\textwidth]{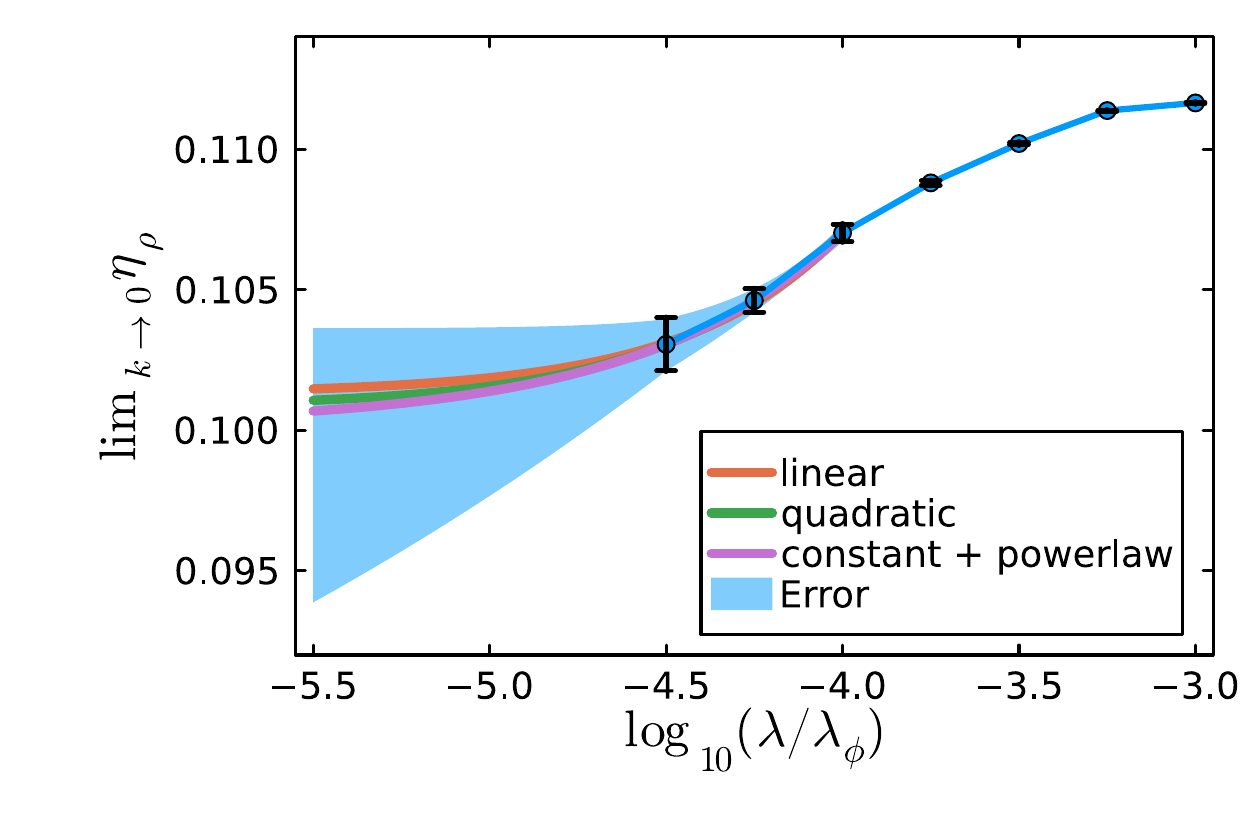}
			\caption{Limit of the Scaling exponent of the propagator spectral function. Several extrapolations with different fit functions are shown. \hspace*{\fill}}
			\label{fig:rhoExtrapolationLimit}
		\end{subfigure}\hfill
		\begin{subfigure}[t]{0.48\linewidth}
			\centering
			\includegraphics[width=\textwidth]{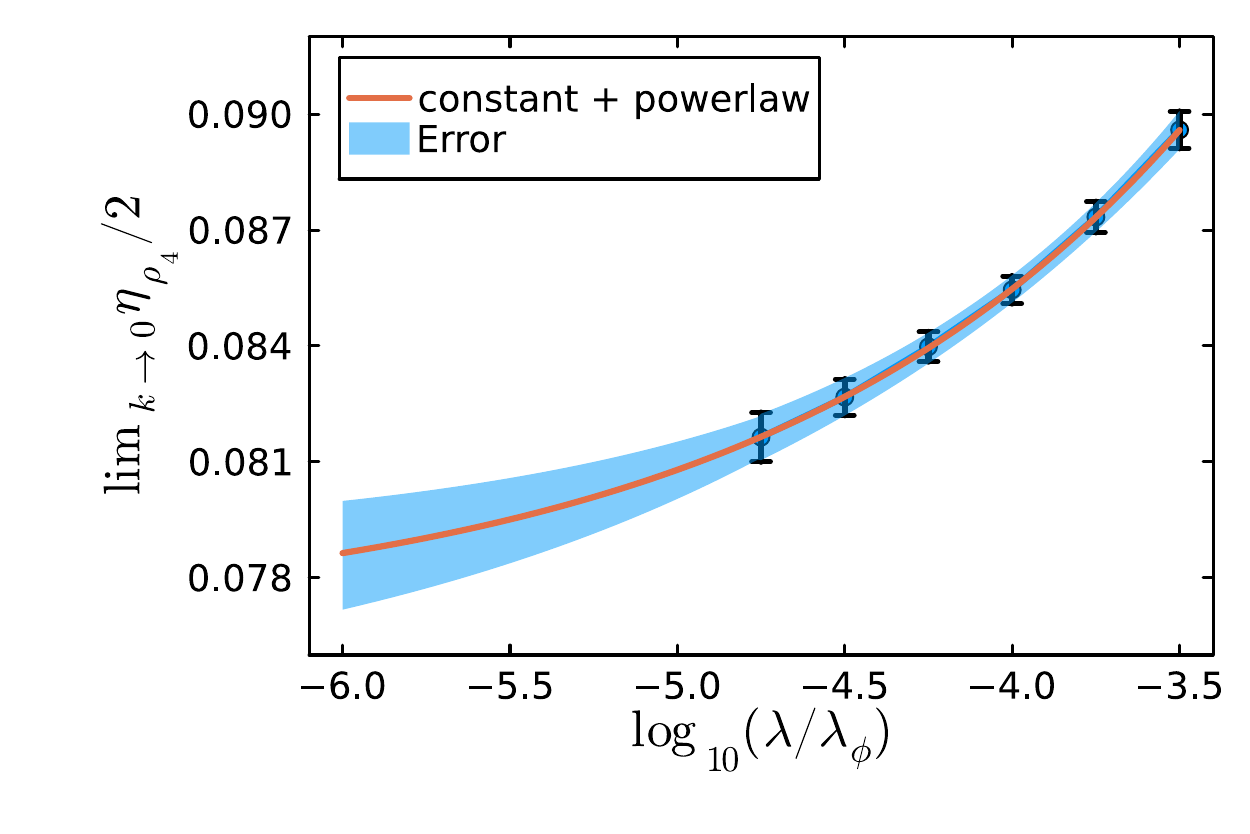}
			\caption{Limit of the Scaling exponent of the four-point spectral function. An extrapolation with a constant plus a power law as fit function is shown. \hspace*{\fill}}
			\label{fig:rho4ExtrapolationLimit}
		\end{subfigure}
		\caption{Limit of the extrapolation $k\to 0$ of the logarithmic derivatives $\eta_\rho$ and $\eta_{\rho_4}$ at different spectral parameters $\lambda$. In~\Cref{fig:Extrapolation} we show explicitly the scaling exponents as a function of $k$. To get the error at $\lambda=0$, we extrapolate as explained in~\Cref{app:ExtrapolationLambda}. The scaling exponent of the propagator only shows signs of convergence in the last three points while the scaling exponent of the four-point function shows convergence throughout. This makes the error of the limit of $\eta_\rho$ significantly larger.
			\hspace*{\fill} } 
		\label{fig:ExtrapolationLimit}
	\end{minipage}
\end{figure*}

To gauge the quality of the extrapolation, we perform it with varying polynomial orders. 
For these fits we extract the limit $k\to 0$, which is given as the constant coefficient of the polynomial, as well as its error, which corresponds to the square root of the first entry in the covariance matrix. 
Additionally, we compute the reduced $\chi$-square value ($\chi^2_{\text{red}}$). 
Both are shown in figure~\Cref{fig:bestPolynomialFit}, exemplary for the point at $\log_{10}(\lambda/\lambda_\phi)=-4.5$ in the propagator spectral function. 
As can be seen in \Cref{fig:bestPolynomialFitLimit}, the extrapolation error of the limit $k\to 0$ of the sliding scale exponent $\eta_\rho$ grows rapidly for polynomial degrees of $N_{\text{max}}>2$. In addition, we observe large fluctuations of the respective coefficient above $N_{\text{max}}\gtrsim 10 $. Together with the very small $\chi^2_{\text{red}}$ value of $\sim 10^{-4}$, these fluctuations are an indication of overfitting. We conclude that a  parabolic fit of the data provides the best results in limit $k\to 0$.
For different spectral parameters, the $N_{\text{max}}$-dependence of the constant coefficient of the polynomial and the $\chi^2_{\text{red}}$ value is very similar: we find $N_{\text{max}}=2$ to be optimal. In contradistinction, the best fit of the values of the logarithmic derivative of the four-point function is given by a fifth-order polynomial.

We show results of the extrapolation with the optimal order of the polynomials in~\Cref{fig:ExtrapolationLimit}. Since we extract the scaling exponent by fitting a slope to an extended area of the logarithmic derivative of the spectral function, we cannot get the limit $k\to 0$ with finer spacing in the spectral parameter $\lambda$ without neighbouring points being correlated.

For lower spectral values in $\rho$, numerical errors in the computation of the spectral functions lead to an increased error in the logarithmic derivative and thus its limit. This can be seen in~\Cref{fig:rhoExtrapolationLimit}, where the error grows with smaller spectral values. For the four-point function, the opposite is the case. It is computed by integrating over the spectral function of the propagator, so it is numerically more stable and the numerical errors depend only mildly on $\lambda$. This is also visible in~\Cref{fig:rho4ExtrapolationLimit}, where all error bars are of approximately the same size.

\subsection{Extrapolation towards $\lambda=0$}
\label{app:ExtrapolationLambda}

Even for the smallest cutoff scale $k$ considered here, the scaling exponents still show a $\lambda$-dependence. Thus, to get a value for the (constant) anomalous dimension, we now have to take the limit $\lambda\to 0$.

The last three points of the limit $k\to 0$ of the scaling exponent $\eta_\rho$ of the propagator spectral function (see~\Cref{fig:rhoExtrapolationLimit}) show the emergence of convergence, and it is expected that the values settle  into a plateau. The plateau value is obtained by an extrapolation. For the respective systematic error estimate we consider three different extrapolations: once with a linear function as fit function, once with a quadratic polynomial and once with a constant plus power law. The three extrapolations are shown in~\Cref{fig:rhoExtrapolationLimit}. We do not embark on a full systematic error determination and simply -roughly- estimate the error by taking the extremal points of the family of curves of these functions which still fit through the error bars. 

To get a value for $\eta_\rho$ we take the average of the extrapolated values. This results in
\begin{align}
	\lim_{\lambda\to 0}\lim_{k\to 0}\eta_\rho = 0.101^{+0.004}_{-0.028}\, .
	\label{eq:etarhoLimit}
\end{align}
For the scaling exponent of the four-point function, we use similar extrapolations. Here, however, the whole range shows convergence. Fitting a constant plus power law shows good agreement with the data. The extrapolation is shown in~\Cref{fig:rho4ExtrapolationLimit} together with an extrapolation error. This results in 
\begin{align}
	\lim_{\lambda\to 0}\lim_{k\to 0}\frac{\eta_{\rho_4}}{2} = 0.077^{+0.002}_{-0.003}\, .
	\label{eq:etarho4Limit}
\end{align}
Since $\rho_4$ is derived from $\rho$ and $Z_\phi$, the scaling of the propagator is the only possible source of scaling for the four-point function. Hence, this value should be interpreted as a mere cross-check, see also the discussion below~\labelcref{eq:eta4Num}. The two values,~\labelcref{eq:etarhoLimit} and~\labelcref{eq:etarho4Limit}, agree within their error bars.

\section{Computation of the LPA${}^\prime$ benchmark}
\label{app:LPAbenchmark}

In this Appendix, we provide some details on the fixed point analysis in LPA${}^\prime$, that is used as a benchmark in~\Cref{sec:LPAprime}.
The flow of the dimensionless effective potential in the three-dimensional scalar theory with a CS-regulator is given as
\begin{align}\nonumber 
	\bigl[ \partial_t -(1+\eta_\phi) \bar\rho\, \partial_{\bar \rho} \bigr]\, u(\bar \rho) +3 \, u(\bar \rho) &\\[1ex]
	&\hspace{-3cm}
	=- \frac{ 2-\eta_\phi}{8 \pi}\, \sqrt{1 +\mu(\bar \rho)} - \dot\mu_\textrm{\tiny{ct}}\, \bar \rho\,, 
	\label{eq:dtukPrime}
\end{align} 
where the $t$-derivative in~\labelcref{eq:dtukPrime} is performed at fixed $\bar\rho$. The field-dependent dimensionless mass function $\mu(\bar \rho)$ in~\labelcref{eq:dtukPrime} reads 
\begin{align} 
	\mu(\bar \rho)= u'(\bar \rho)+2\bar{\rho}\,u''(\bar \rho)\,.
	\label{eq:omegaLPADefinition}
\end{align}
The dimensionless potential $u$ and field $\bar\rho$ are defined by rescaling the dimensionful quantities by appropriate powers of the cutoff scale $k$, 
\begin{align} 
	\hspace{-.2cm}u(\bar\rho) = v(\bar \rho)- \bar \rho\,, \quad v(\bar \rho)=\frac{V_\textrm{eff}(\rho)}{k^3} \,, \quad
	   \bar\rho = Z_\phi \,\frac{ \rho}{k}\,. 
	\label{eq:Dimlessurho}
\end{align}
\Cref{eq:Dimlessurho} follows the standard fRG convention: the cutoff term is subtracted from the effective action. Keeping this standard convention allows for a straightforward comparison. 

The term $-\dot\mu_\textrm{\tiny{ct}} \, \bar \rho$ is a part of the flow $\partial_t S_\textrm{\tiny{ct}}$ of the counter term action in the renormalised CS-equation~\labelcref{eq:CSflow}. The rest of $\partial_t S_\textrm{\tiny{ct}}$ ensures the manifest finiteness of the renormalised CS-equation, see \cite{Braun:2022mgx}, while the choice of $\dot\mu_\textrm{\tiny{ct}} $ adjusts for the renormalisation condition, 
\begin{align}
\partial_t S_\textrm{\tiny{ct}} \simeq  	\dot\mu_\textrm{\tiny{ct}}\,\int\! \mathrm{d}^3x  \bar \rho\,.  
\label{eq:SctLPAprime}	 
\end{align} 
Note also that the standard fixed point analysis uses the CS-regulator $R_\textrm{\tiny{CS}} = +Z_\phi k^2$ in the symmetric phase, despite the fixed point potential being in the broken phase. Finiteness in the three-dimensional theory is achieved by studying the flow of the $\bar \rho$-derivative of $u$, that is $\partial_t u'$. This translates into a flowing counter term action with $\dot\mu_\textrm{\tiny{ct}}=0$ in~\labelcref{eq:dtukPrime}. We note in passing, that $\dot\mu_\textrm{\tiny{ct}}$ also accommodates flowing renormalisation conditions for momentum-dependent regulators, see \cite{Braun:2022mgx} which can be used for an optimisation of the convergence properties of a given expansion.

\begin{table}[t]
	\centering
	\begin{tabular}{|c|c|c|c|c|c|}
		\hline
		$n$        & 2       & 4       & 6       & 8       & 9       \rule[-1.3ex]{0pt}{4ex} \\
		\hline
		$\eta$     & \num{0.1600} & \num{0.0829} & \num{0.0804} & \num{0.0802} & \num{0.0802} \rule[-1.3ex]{0pt}{4ex} \\
		\hline
	\end{tabular}
	\caption{Anomalous dimension $\eta$ from the LPA${}^\prime$ computation with the CS-cutoff for different maximal power $\bar\rho^n$ in the effective potential $u$.~\labelcref{eq:Dimlessurho}.\hspace*{\fill}}
	\label{tab:LPAResults}
\end{table}

The fixed point equation for the effective potential follows from~\labelcref{eq:dtukPrime} with $\partial_t u \equiv 0$. Note that the split~\labelcref{eq:Dimlessurho} into potential and regulator contribution is necessary for this fixed point condition. The full dimensionless effective potential $v(\bar \rho)$ includes the trivial running of the regulator term $\bar \rho$. The corresponding fixed point equation reads $\partial_t v(\bar \rho)= -(1+\eta)\bar\rho $. For the iterative solution of the fixed point equation, we use the Taylor series representation of $u$ within an expansion about the $t$-dependent minimum $\kappa$, 
\begin{align}
	u(\bar{\rho})=\sum_{n\geq 2}\frac{\lambda_n}{n!}(\bar{\rho}-\kappa)^n\, , 
\label{eq:uexpansion}
\end{align}
valid for $\kappa> 0$ with 
\begin{align}  
	\left. \frac{\partial u(\bar \rho)}{\partial \bar\rho}\right|_{\bar\rho =\kappa}  =0\,. 
	\label{eq:EoMu}
\end{align}
The $\lambda_n$ are the dimensionless and RG-invariant scattering couplings of the $\phi^4$-theory. The classical potential is given by~\labelcref{eq:uexpansion} with 
\begin{align} 
	\lambda_2 = \frac23 \lambda_\phi\,,\qquad \lambda_{n\neq 2}=0\,.
\label{eq:Classicalu}
\end{align} 
The critical scaling on the phase transition is obtained by evaluating~\labelcref{eq:dtukPrime} on the fixed point $u^*$ with  
\begin{align}
	\partial_t u^*(\bar\rho)=0\,.
	\label{eq:u*}
\end{align}
This leads us to the fixed point equation 
\begin{align} \nonumber 
3\, u^*(\bar \rho)  - (1+\eta_\phi) \bar\rho\,  {u^*}'(\bar \rho) & \\[1ex] 
&\hspace{-2cm} =- \frac{ 2-\eta_\phi}{8 \pi}\, \sqrt{s_\pm +\mu^*(\bar \rho)}-\dot\mu_\textrm{\tiny{ct}}  \bar \rho\,.
\label{eq:FPu}
	\end{align}
This equation is complemented by the algebraic relation for $\eta_\phi$, 
\begin{align}
	\eta_\phi =&\, \frac{2\lambda_2\mu_0}{\lambda_2\mu_0+8\pi a_\eta}\,, \\[2ex] \nonumber
	a_\eta=&\,\left(4+4\sqrt{1+\mu_0}+\mu_0\left(5+\mu_0+3\sqrt{1+\mu_0}\right)\right)\,,
	\label{eq:etaphiRen0}
\end{align}
with $\mu_0= \mu(\kappa)$ in~\labelcref{eq:omegaLPADefinition}, evaluated on the solution $\kappa$ of the equation of motion~\labelcref{eq:EoMu}.

We have computed the anomalous dimensions $\eta$ on the fixed point for different truncations of the effective potential. The truncation best suited for comparison with the spectral flow is one where we only used the terms up to $\bar{\rho}^2$ in the effective potential. To analyse the improvement of the results with higher orders, we went up to $\bar{\rho}^9$, see~\Cref{tab:LPAResults}. The results are discussed in~\Cref{sec:LPAprime}.

\section{Technical details and numerics}
\label{app:Numerics}

In this Appendix, we provide technical details on the numerical solution of the spectral flow equation.

\subsection{Numerical implementation}

The numerical implementation uses Julia~\cite{Julia-2017}. The two-point function (and therefore the spectral function) was calculated on a logarithmic momentum grid, which over the course of the flow was expanded to include lower momenta. To interpolate the values between the grid points, we used a third order spline on the logarithm of the computed values above the onset. The integrals over the spectral functions were solved using Gaussian quadrature with a logarithmic substitution, because the prominent features of the spectral functions were several orders of magnitude smaller than the integration region. Integrable singularities of principal value integrals were subtracted and solved analytically.

We solved the flow equation~\labelcref{eq:CSflow}, more specifically~\labelcref{eq:FlowG2Final}, using a Runge-Kutta method of third order with a step size $\Delta\log_{10}(k/\lambda_\phi)=0.005$. Flowing in $\log_{10}(k)$ allowed us to keep a constant step size over the whole flow.

\subsection{Initial condition}
\label{app:initial conditions}

At a sufficiently high renormalisation scale $\Lambda$, we can approximate $\Gamma^{(2)}$ using perturbation theory. The first relevant diagram is the sunset diagram with classical propagators and vertices at two-loop-level, since the constant tadpole is absorbed in the renormalisation condition. Therefore, the initial condition for the two-point function $\Gamma_\Lambda^{(2)}$ is
\begin{align}
	\Gamma_\Lambda^{(2)}(\omega) = m_{\text{\tiny{pole}}}^2-\omega^2 - \frac{\lambda_\phi^2}{6} \left[D_{\text{sun}}(\omega)-D_{\text{sun}}(m_{\text{\tiny{pole}}})\right]\,,
\end{align}
with the sunset diagram on two-loop level
\begin{align}\nonumber
	D_{\text{sun}}(\omega) = &\frac{1}{Z_{\phi,\Lambda}^3}\frac{1}{(4\pi)^2}\Bigg[\frac{1}{2}\log\left(\frac{1}{(3m_{\text{\tiny{pole}}})^2-\omega^2}\right)\\
	&-\frac{3m_{\text{\tiny{pole}}}}{\omega}\text{arctanh}\left(\frac{\omega}{3m_{\text{\tiny{pole}}}}\right)\Bigg]
\end{align}
This formula is taken from~\cite{Horak:2020eng}. To achieve RG-consistency, and hence avoid potential cusps or discontinuities in the solution, we use an RG-improvement by computing the initial residue $\frac{1}{Z_{\phi,\Lambda}}$ iteratively from the sum rule.

\subsection{Sum rule}
\label{app:NumericsFlow}

The frequency dependent two point function can be written as 
\begin{align}
	\Gamma^{(2)}(\omega) & \equiv k^2-\omega^2+\Pi_k(\omega)\,,
\end{align}
where $\Pi_k(\omega)$ is the loop induced self energy. 
In the deep IR, the finite numerical precision leads to small, numerical deviations from the sum rule, which, if not corrected, can build up to destabilize the flow. To ensure that the sum rule is always fulfilled, we rescale the spectral tail of the propagator by a factor
\begin{align}
	r=\frac{1-\frac{1}{Z_\phi} }{\int_\lambda\tilde{\rho}(\lambda)}\,.
\end{align}
This leads to a change in $\Pi_k$ by
\begin{align}
	\Pi_k(\omega)\to\left(\frac{1}{r}-1\right)(k^2-\omega^2)+\frac{1}{r}\Pi_k(\omega).
\end{align}
%

\section{Computation of diagrams}
\label{app:analyticintegrands}

In this Appendix, we derive the analytic expressions for the diagrams~\labelcref{eq:Ddyntad,eq:Dfish}, resulting from the momentum integrations. More details concerning these computations, in particular the finiteness and commutativity of the momentum and spectral integrations can be found in \cite{Horak:2020eng,Braun:2022mgx, Eichmann:2023tjk, Horak:2023hkp}.

\subsection{Spectral diagrams}
\label{app:specDiags}

Performing the momentum integral in the tadpole diagram~\labelcref{eq:Ddyntad} leads to  
\begin{subequations} 
	\label{eq:DdyntadApp}
\begin{align}
\hspace{-.1cm}	D_{\text{tad}}^{\text{dyn}}(p) =\int\limits_{\lambda}\, \rho(\lambda_1)\rho(\lambda_2)\rho_4(\lambda_3)I_{\text{pol}}(\lambda_1,\lambda_2,\lambda_3,p)\,.
\label{eq:DdyntadApp1}
\end{align}
with the analytic result of the momentum integration in  
\begin{align}\nonumber
\hspace{-.2cm}	I_{\text{pol}}(\lambda_1,\lambda_2,\lambda_3,p) =&\, \int\limits_q\! \frac{1}{(\lambda_1^2+q^2)(\lambda_2^2+q^2)(\lambda_3^2+(q+p)^2)} \\*[2ex]
	 &\hspace{-2.7cm}= \frac{-1}{\lambda_1^2-\lambda_2^2}\left[\Tilde{I}_{\text{pol}}(\lambda_1,\lambda_3,p)-\Tilde{I}_{\text{pol}}(\lambda_2,\lambda_3,p)\right]\, .
	\label{eq:Ipol}
\end{align}
and
\begin{align}\nonumber
	\Tilde{I}_{\text{pol}}(\lambda_1,\lambda_2,p)      =&\, \int\limits_q\,  \frac{1}{(\lambda_1^2+q^2)(\lambda_2^2+(q+p)^2)}  \\*[2ex]
	\Tilde{I}_{\text{pol}}(\lambda_1,\lambda_2,\omega) =& \,\frac{1}{4\pi \omega}\, \text{arctanh}\left(\frac{\omega}{\lambda_1+\lambda_2}
	\right)\,.     
	\label{eq:tildeIpol}
\end{align}
\end{subequations}
The $\omega$ dependence in the last line indicates the evaluation on the real frequency axis. Note, that we use the convention $\Im(\text{arctanh}(x>1))=+\frac{\pi}{2}$, which corresponds to approaching the branch cut from above, i.e., the retarded limit $p_0\to -\imag (\omega+\imag 0^+)$.
Performing the momentum integral in the fish diagram~\labelcref{eq:Dfish}, we arrive at
\begin{align}\nonumber 
 D_{\textrm{\tiny{fish}}}(p)    =& \int\frac{\textrm{d}^3q}{(2\pi)^3}G(q)G(q+p) \\[2ex]
  =&\int\limits_{\lambda}\, \rho(\lambda_1)\rho(\lambda_2)\tilde{I}_{\text{pol}}(\lambda_1,\lambda_2,p)\,, 
\label{eq:DfishApp}
\end{align}
with $\tilde{I}_{\text{pol}}$ in~\labelcref{eq:tildeIpol}. Crucially, the analytic dependence of $\tilde{I}_{\text{pol}}$ on $p$ allows us to evaluate all diagrams in the full complex frequency plane. All non-perturbative information about the propagators is then accounted for by the spectral integrals, which are computed numerically.

\subsection{Speeding up the numerical computation of the spectral integrals}
\label{app:SpeedUpSectral}

While the representation of the tadpole diagram in~\labelcref{eq:DdyntadApp} is perfectly valid, the numerical evaluation of the spectral integrals is the bottleneck of our computation. It is computationally convenient to use a spectral representation of the squared propagator,
\begin{subequations}
\begin{align}
	G^2(p) = \int\limits_{\lambda}\frac{\rho_2(\lambda)}{p^2+\lambda^2}\,,
\end{align}
which is guaranteed to exist, if the spectral representation of the propagator does. Given the propagator spectral function~\labelcref{eq:rhoparam}, the corresponding spectral function of the squared propagator reads
\begin{align}\label{eq:rho2}
	\rho_2(\omega) & =2\Im[G^2(p=i(\omega+i0^+))]     \\*\nonumber
	               & =\frac{2\pi}{Z_\phi^2}\partial_{\omega^2}\delta(m_{\text{\tiny{pole}}}^2-\omega^2)  \\*\nonumber 
	               & \quad+\frac{4\pi}{Z_\phi}\delta(m_{\text{\tiny{pole}}}^2-\omega^2)\int_{m_{\text{scat}}}^\infty\textrm{d}\lambda\ \frac{\lambda}{\pi}\frac{\Tilde{\rho}(\lambda)}{\lambda^2-m_{\text{\tiny{pole}}}^2}                              \\*\nonumber
	               & \quad-\frac{2}{Z_\phi}\frac{\Tilde{\rho}(\omega)}{\omega^2-m_{\text{\tiny{pole}}}^2}-\int_{m_{\text{scat}}}^\infty\textrm{d}\lambda\ \frac{2\lambda}{\pi}\frac{\Tilde{\rho}(\lambda)\Tilde{\rho}(\omega)}{\omega^2-\lambda^2}.
\end{align}
\end{subequations}
The last two terms contain no delta poles, and we refer to them as $\tilde{\rho}_2(\omega)$. 
This reduces the maximal dimensionality of the spectral integrands by one. The respective expression for the tadpole diagram reads 
\begin{align}
	D_{\text{tad}}^{\text{dyn}}(p) &= \int\limits_{\lambda}\, \rho_2(\lambda_1)\rho_4(\lambda_2)\Tilde{I}_{\text{pol}}(\lambda_1,\lambda_2,p)\,.
\end{align}
The remaining two-dimensional integrals can be further simplified using convolutions. For that, we observe that $\tilde{I}_{\text{pol}}$ only depends on $\lambda_1+\lambda_2$ and $\omega$. This allows for the following substitution:
\begin{align}\nonumber
	 & \int\textrm{d}\lambda_1\textrm{d}\lambda_2\ \frac{\lambda_1\lambda_2}{\pi^2}\rho_i(\lambda_1)\rho_j(\lambda_2)\Tilde{I}_{pol}(\lambda_1,\lambda_2, p)         \\
	 & =\int\textrm{d}\lambda_1\textrm{d}\nu\ \frac{\lambda_1(\nu-\lambda_1)}{\pi^2}\rho_i(\lambda_1)\rho_j(\nu-\lambda_1)\Tilde{I}_{\text{pol}}(\nu, 0, p)\nonumber \\
	 & =\int\frac{\textrm{d}\nu}{\pi^2}\ (\rho_i'*\rho_j')(\nu)\Tilde{I}_{\text{pol}}(\nu, 0, p)
\end{align}
with $\rho_i'(\lambda)=\lambda\rho_i(\lambda)$. This reduces the numerical complexity of the two-dimensional integral to that of two consecutive one-dimensional ones on separate one-dimensional grids. In doing so, it arranges for the non-analyticities to be resolved in only one dimension, significantly reducing the number of numerical evaluations.

\section{Systematic improvements}
\label{app:fullpotential}

In this Appendix, we discuss two straightforward systematic improvements of the present approximation: the first improvement concerns the effective potential, that has only been considered here in a low order of the Taylor expansion. The full effective potential is readily included by a straightforward extension of the present approximation, using computational methods that have been already set up and tested in similar situations. This is discussed in~\Cref{app:FullVeff}. The second improvement concerns the momentum dependence of the four-point function. In the present work, we used the $s$-channel resummation, and lifting it is discussed in~\Cref{app:FullMomentum4point}. Apart from the quantitative improvement, this is also relevant for obtaining the scaling laws for all momenta and spectral values on the fixed point.

\subsection{Full effective potential and the broken regime}
\label{app:FullVeff}

To achieve quantitative precision for scaling exponents, the resolution of higher order scatterings is crucial, as discussed in detail in~\Cref{sec:LPAprime}. This can be accommodated for by coupling the flow of the propagator to that of the effective potential. The latter can be treated either by a converging order of the Taylor expansion around the minimum or by solving the full partial differential equation for the effective potential. An iterative scheme for solving the coupled set of momentum-dependent equations and the flow of the full effective potential has been set up in \cite{Helmboldt:2014iya}. Apart from this simple scheme we aim to use a comprehensive scheme that builds on the numerics advances reported on in \cite{Grossi:2019urj, Grossi:2021ksl, Koenigstein:2021syz, Koenigstein:2021rxj, Ihssen:2022xkr, Ihssen:2023qaq, Ihssen:2023xlp,Ihssen:2024miv, Zorbach:2024rre}. Finally, we will extend the current analysis to the broken regime, see \cite{Horak:2023hkp, Eichmann:2023tjk}. 
We hope to report on respective results in the near future.

\subsection{Momentum dependence of the four-point function}
\label{app:FullMomentum4point}

The four-point function only appears in the tadpole-diagram, which only depends on the momentum configuration $\Gamma^{(4)}(p,q,-q,-p)$. Here, $p$ is the external momentum and $q$ the loop momentum. In this specific momentum configuration, the $s$- and $u$-channels contribute equally. The $t$-channel is evaluated at $t=0$ and only contributes an additive constant in such a configuration. Accordingly, $\Gamma^{(4)}(p,q,-q,-p)$ only depends on $s,u$ and the angle $\theta$ with $\cos \theta = p_\mu q_\mu /\sqrt{s\, t}$. This angle is integrated over in the loop integration and hence the angular dependence is averaged out, for a respective analysis for four-quark scatterings QCD see \cite{Fu:2025hcm}. Consequently, we can consider this dependence to be subleading and, in a first step, resort to a two-channel approximation with $\Gamma^{(4)}(p,q,-q,-p)\to \Gamma^{(4)}(s,t)$. 

The inclusion of $\Gamma^{(4)}(s,t)$ improves the quantitative precision in comparison to the present computation. Moreover, the constant contribution of the channels shifts the intrinsic scale of the four-point function to higher momenta. To illustrate this, we consider our $s$-channel approximation. Its UV-limit is given by the classical coupling, as the loop integral decays. To accommodate for the $t$-channel contribution in the resummation, we have to replace the classical coupling in~\labelcref{eq:G4Dyn} by an effective coupling $\lambda_{\text{\tiny{eff}}}=\lambda_\phi + \int \frac{dk}{k} \textrm{flow}_{\text{\tiny{const}}}$. The second term comprises the diagrams of the flow of the 4-point function, which do not depend on the $s$-channel momentum. These terms in the flow are proportional to $\lambda_\phi^2$. The dimensionality of the four-point function hence requires that they are roughly proportional to $1/k$ which is the only other scale in the theory. This shifts the UV-boundary of the scaling region to higher momenta when integrating towards the phase boundary at $k=0$. In combination with the fine-tuning of the initial condition, this potentially removes all intrinsic scales in the computation and allows to approach the uniform scaling limit. We shall report on these advances in the near future.


\bibliography{spectralfrgDraft.bib}

\end{document}